\documentclass[10pt]{article}
\usepackage{geometry}
\usepackage{xcolor}
\definecolor{graycolor}{gray}{0.9} 
\usepackage{microtype}
\usepackage{setspace} 
\usepackage[utf8]{inputenc}
\usepackage[english]{babel}
\usepackage{times}
\usepackage{array}
\usepackage{soul}
\usepackage{cite}
\usepackage[numbers,sort&compress]{natbib}
\usepackage{natbib}

\usepackage{titlesec} 
\titleformat {\section} [block] {\raggedright \fontsize{10}{10}\selectfont\bfseries} {\thesection. \space} {0pt} {}
\titlespacing {\section} {0pt} {12pt} {6pt}
\titleformat {\subsection} [block] {\raggedright \fontsize{10}{10}\selectfont\itshape} {\thesubsection .\space} {0pt} {}
\titlespacing {\subsection} {0pt} {12pt} {6pt}
\titleformat {\subsubsection} [block] {\raggedright \fontsize{10}{10}\selectfont} {\thesubsubsection .\space} {0pt} {}
\titlespacing {\subsubsection} {0pt} {12pt} {6pt}
\titleformat {\paragraph} [block] {\raggedright \fontsize{10}{10}\selectfont} {} {0pt} {}
\titlespacing {\paragraph} {0pt} {12pt} {6pt}

\usepackage{array} \newcommand{\PreserveBackslash}[1]{\let\temp=\\#1\let\\=\temp}
\newcolumntype{C}[1]{>{\PreserveBackslash\centering}m{#1}}
\newcolumntype{R}[1]{>{\PreserveBackslash\raggedleft}m{#1}}
\newcolumntype{L}[1]{>{\PreserveBackslash\raggedright}m{#1}}
\usepackage{lineno}
\usepackage{tabularx}
\usepackage{colortbl}
\usepackage{graphicx}
\usepackage{float}
\usepackage[export]{adjustbox}
\usepackage{caption}
\usepackage{fancyhdr} 
\usepackage{amsmath}
\usepackage{bm}
\setcitestyle{open={[},close={]},citesep={,\!},numbers}
\usepackage{lastpage}
\usepackage{layout}
\usepackage{setspace} 
\usepackage{enumitem}
\usepackage{booktabs}
\usepackage{arydshln}
\usepackage{multirow}
\usepackage{color}
\usepackage{hyperref} 
\hypersetup{
	colorlinks=true,
	linkcolor=blue,
	filecolor=blue,
	urlcolor=black,
	citecolor=cyan,
}

\fancypagestyle{firstpage}{
    \setlength{\headsep}{2.2cm}
    
    \setlength{\footskip}{1.5cm}
    \fancyhf{}
    \lhead{\begin{table}[H]
        \centering
        \begin{tabular}{L{2.5cm}C{10cm}C{3.1cm}R{2cm}}
            \includegraphics[scale=0.035]{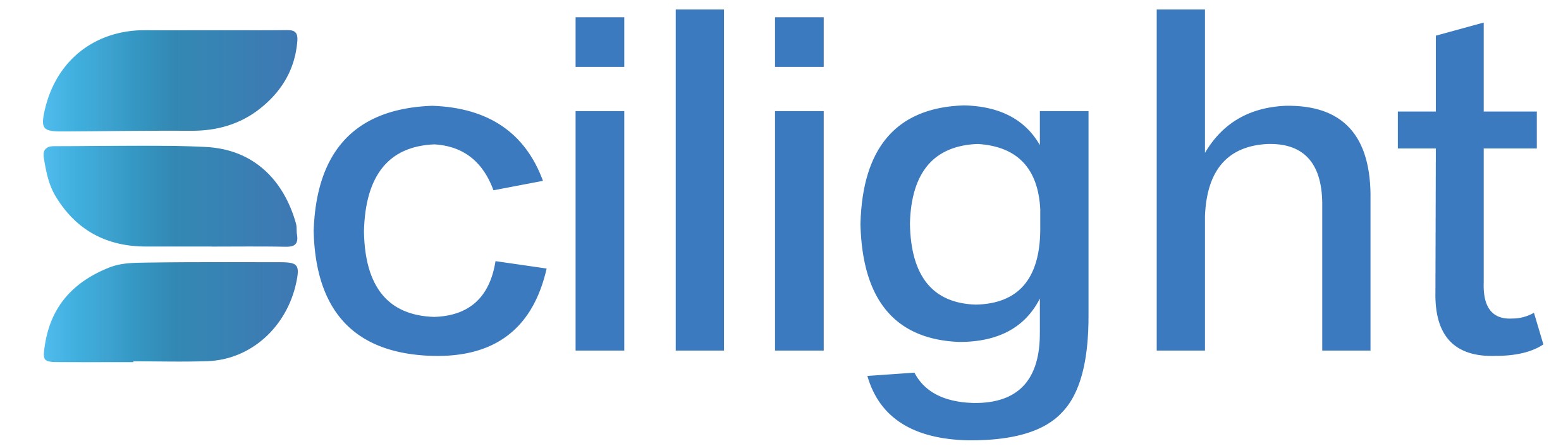} \vspace{-6pt}& \cellcolor{graycolor}\begin{tabular}[c]{@{}c@{}}\textit{Physics and the Cosmos}\\ \href{https://www.sciltp.com/journals/pac}{https://www.sciltp.com/journals/pac}\end{tabular} & \includegraphics[scale=0.022]{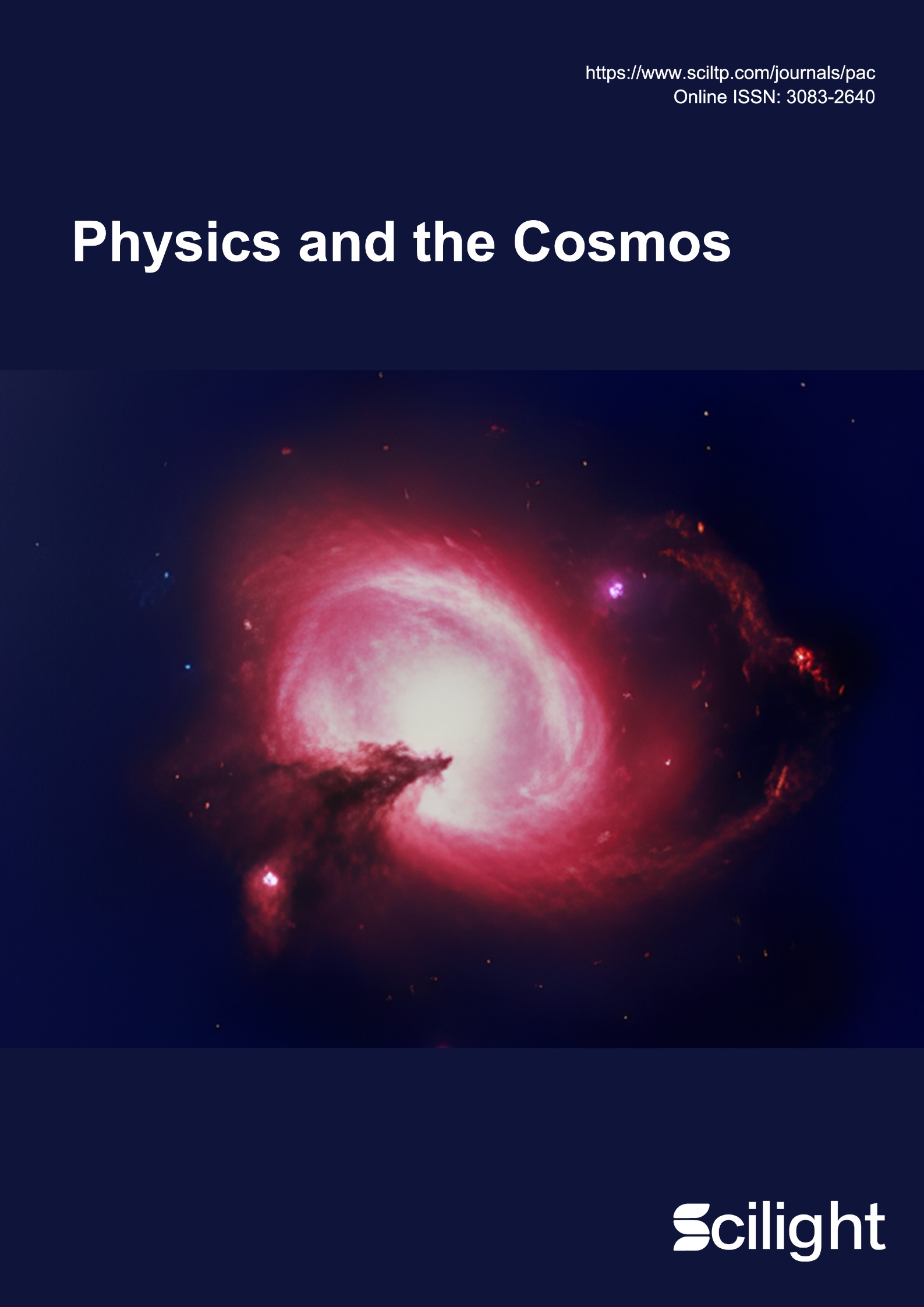} \vspace{-3pt}\\
        \end{tabular}
        \vspace{-22pt}
    \end{table}}
   
    \fancyfoot[C]{
        \vspace{-1.55cm}
        \begin{table}[H]
            \begin{minipage}[c]{0.15\columnwidth}
                \includegraphics[scale=0.5]{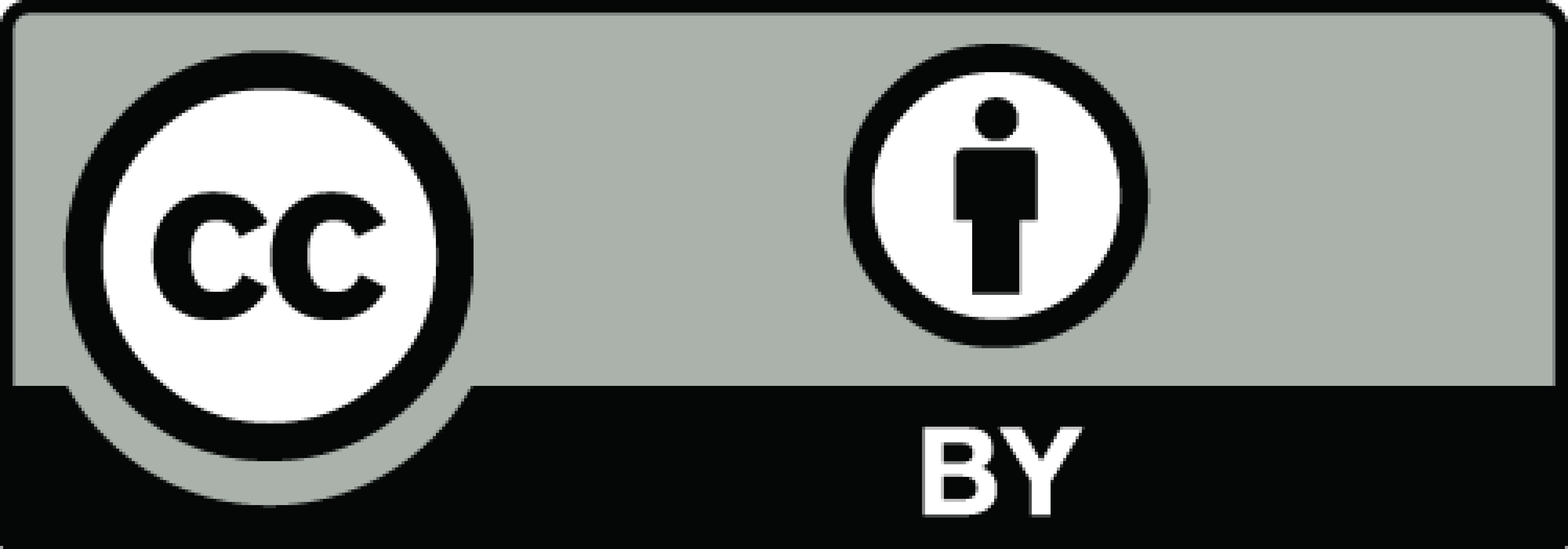} \vspace{1.1pt}
            \end{minipage}
            \hfill
            \begin{minipage}[c]{0.85\columnwidth}
                \scriptsize \textbf{Copyright:} © 2026 by the authors. This is an open access article under the terms and conditions of the Creative Commons Attribution (\mbox{CC BY}) license  (\href{https://creativecommons.org/licenses/by/4.0/}{https://creativecommons.org/licenses/by/4.0/}). \\ \textbf{Publisher’s Note:} Scilight stays neutral with regard to jurisdictional claims in published maps and institutional affiliations.
            \end{minipage}
    \end{table}}
    \vspace{-0.55cm}
}

\begin{document}
\newgeometry{left=2.5cm, right=2.5cm, top=1.8cm, bottom=4cm}
	\thispagestyle{firstpage}
	\nolinenumbers
	{\noindent \textit{Review}}
	\vspace{4pt} \\
	{\fontsize{18pt}{10pt}\textbf{The Quest for Neutrinoless Double Beta Decay: Progress \\and Prospects}}
	\vspace{16pt} \\
	{\large Andrea Giuliani}
	\vspace{6pt}
	 \begin{spacing}{0.9}
		{\noindent \small
			\parbox[t]{0.98\linewidth}{{IJCLab, CNRS/IN2P3, Universit\'e Paris-Saclay}
, 91405 Orsay, France; andrea.giuliani@ijclab.in2p3.fr; Tel.: +33-1-69155526}
			\vspace{6pt}\\
		\footnotesize	\textbf{How To Cite}: Giuliani, A. The Quest for Neutrinoless Double Beta Decay: Progress and Prospects. \emph{Physics and the Cosmos} \textbf{2026}, \emph{1}(1), 6. \href{https://doi.org/10.53941/pac.2026.100006}{https://doi.org/10.53941/pac.2026.100006}}\\
	\end{spacing}

\begin{table}[H]
\noindent\rule[0.15\baselineskip]{\textwidth}{0.5pt} 
\begin{tabular}{lp{12cm}}  
 \small 
  \begin{tabular}[t]{@{}l@{}} 
  \footnotesize  Received: 18 January 2026 \\
  \footnotesize  Revised: 28 February 2026 \\
   \footnotesize Accepted: 3 March 2026 \\
  \footnotesize  Published: 24 March 2026
  \end{tabular} &
  \textbf{Abstract:} Neutrinoless double beta decay is a hypothetical nuclear transition whose observation would demonstrate that neutrinos are their own antiparticles and that lepton number is not conserved, with far-reaching implications for the origin of neutrino mass and the matter–antimatter imbalance in the Universe. This review examines the theoretical foundations of this process and surveys the principal experimental strategies developed to search for it, focusing on their operating concepts, strengths, and limitations. We summarize the current experimental landscape by presenting the most sensitive results achieved so far and by outlining the complementary approaches pursued by different detection techniques. Finally, we discuss the future direction of the field, emphasizing the technological advances needed to reach substantially better sensitivities and, ultimately, to detect this rare phenomenon. \\
\\
  & 
  \textbf{Keywords:} lepton number violation; neutrino nature; neutrino mass; rare decays 
\end{tabular}
\noindent\rule[0.15\baselineskip]{\textwidth}{0.5pt} 
\end{table}

	\section{Introduction }
    \label{sec:intro}

	Neutrinos, subatomic particles with weak interactions and long assumed to be massless, challenge the Standard Model (SM) of particle physics. The discovery of flavor oscillations, which proves that neutrinos have mass, contradicts this SM assumption and opens new directions in fundamental physics. The existence of massive neutrinos opens the possibility of neutrinoless double beta decay ($0\nu2\beta$), a yet-undetected very rare process in which an even-even nucleus $(A,Z)$ decays to an isobar $(A,Z+2)$, emitting two electrons and no other particles. Observing $0\nu2\beta$ would reveal the fundamental nature of neutrinos, confirming that they are Majorana particles: fermions identical to their own antiparticles. Such a discovery would help explain the smallness of neutrino masses and contribute to our understanding of the Universe's matter–antimatter asymmetry. Moreover, $0\nu2\beta$ provides a test of lepton-number conservation, an accidental SM symmetry expected to be violated in many of its extensions. 
    
    In this article, we introduce the basic concepts related to $0\nu2\beta$ and discuss their profound implications for fundamental particle physics and for cosmology. We then turn to experimental methods: after examining the most sensitive detection strategies, along with their specific advantages and drawbacks, we review the results of the most advanced current experiments. Finally, we attempt to provide a long-term perspective on the study of this process, considering the scale and nature of the efforts required to extend half-life sensitivity to the $10^{28}$--$10^{30}$\,yr range, which may be necessary to achieve an actual detection of this elusive and fascinating process.

    Numerous reviews on double beta decay have been published over the past decades. We cite here two recent and especially comprehensive ones~\cite{Agostini:2022zub,Gomez-Cadenas:2023}.
    
	\section{Double Beta Decay: Basic Concepts}
    \label{sec:concepts}

\begin{figure}[H]
    \centering
    \includegraphics[width=0.9\textwidth]{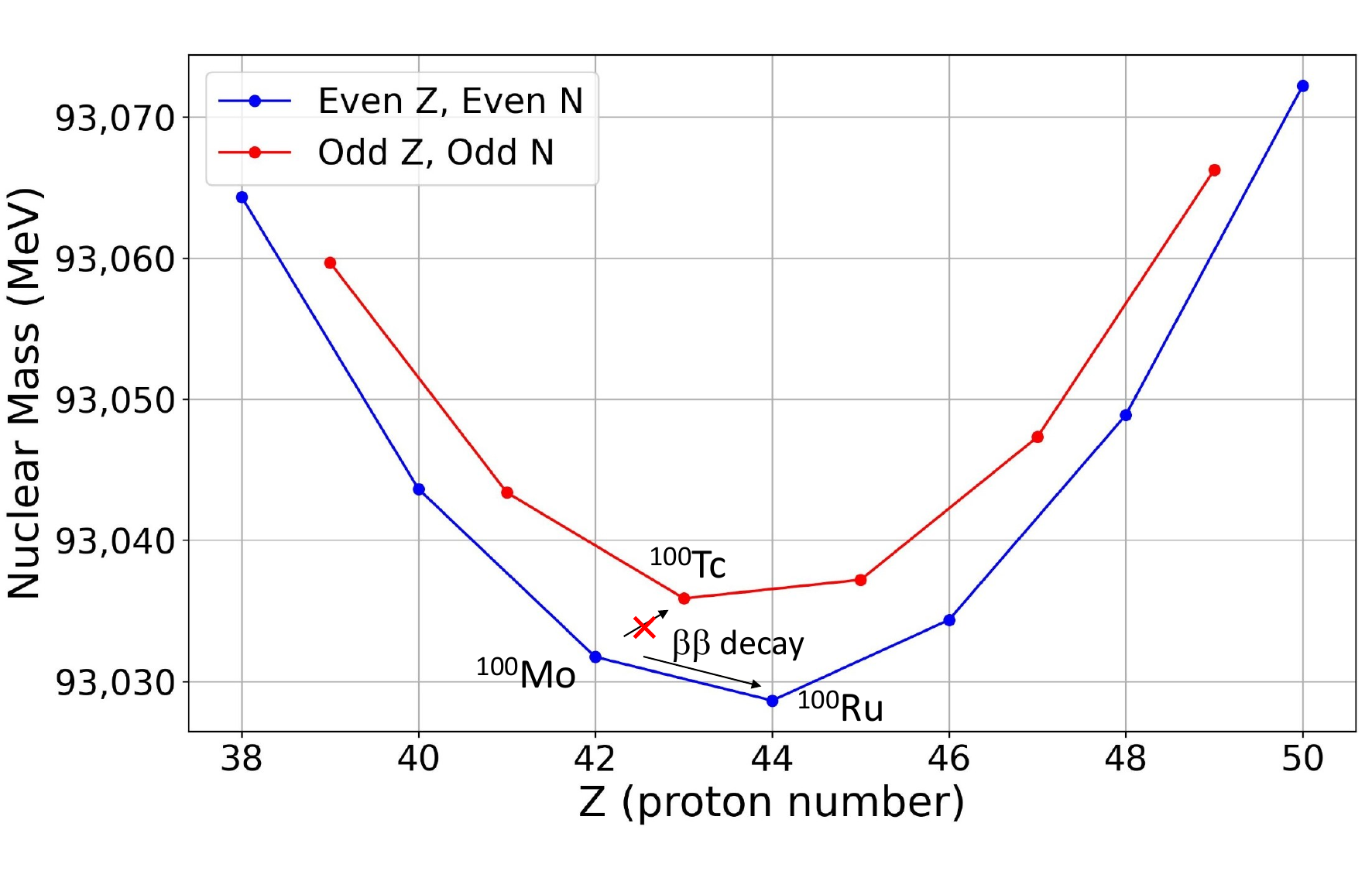}
\vspace{-16pt}
    \caption{{Nuclear mass parabolae for A = 100 according to the Weisz\"acker Formula. Single beta decay of $^{100}$Mo cannot proceed because of energy conservation, while double beta decay is possible.} 
}
    \label{fig:mass-parabolae}
    \end{figure}

    Double beta decay is the rarest process governed by the nuclear weak interaction. It occurs in even-even nuclei for which the single beta transition to the intermediate odd-odd system is energetically inaccessible or severely hindered by a large change of the nuclear spin-parity state. This feature arises from nuclear pairing, which lowers the masses of even-even isotopes relative to their odd-odd neighbors within the same isobaric chain.
 Only because of this pairing-induced energy pattern can double beta decay proceed, with a metastable isobar changing into a lighter one by the simultaneous emission of two electrons, as illustrated in Figure~\ref{fig:mass-parabolae}. In a broad sense, double beta 
\restoregeometry

\noindent decay also includes the  $\beta^+\beta^+$, $\beta^+\mathrm{EC}$, and $\mathrm{EC}\mathrm{EC}$ modes, corresponding to transitions on the right-hand side of the mass parabolae in Figure~\ref{fig:mass-parabolae}. There are 35 nuclei that can undergo double beta decay. When restricting the discussion to the $\beta^-\beta^-$ channel---which is experimentally the most attractive owing to its larger atomic transition energies---the number of candidate nuclides is reduced to 22. In the following, we restrict the discussion to the $\beta^-\beta^-$ channel connecting the $0^+$ ground states of the parent and daughter nuclei. Transitions to excited states are also possible, but they are characterized by significantly reduced decay probabilities as well, due to the smaller available transition energy. 
\vspace{-12pt}

       \vspace{-15pt} 

    The two-neutrino mode ($2\nu2\beta$), which involves the emission of two electron antineutrinos together with the two electrons (respecting lepton-number conservation and the SM principles), was first suggested by Goeppert-Mayer in 1935~\cite{Goeppert-Mayer:1935}. As a second-order weak process, its extremely small probability explains the extraordinarily long lifetimes involved; it was not directly observed until 1987~\cite{Moe:1987}. Since that discovery, the process has been identified in a dozen isotopes, exhibiting half-lives ranging from about $10^{18}$ to $10^{24}$~yr~\cite{Barabash:2020}. A distinct possibility is the neutrinoless channel, first introduced by Furry~\cite{Furry:1939} following the proposal by Majorana that the neutrino might be its own antiparticle~\cite{Majorana:1937}, implying lepton number violation with $\Delta L = 2$. This variant lies therefore outside the minimal electroweak theory. 
    
    The nuclear transitions for $2\nu2\beta$ and $0\nu2\beta$ are respectively:
    \begin{align}
    {}_{Z}^{A}X_{N} &\;\rightarrow\; {}_{Z+2}^{A}X_{N-2} + 2e^{-} + 2\overline{\nu}_{e}, \label{eq:2nubb_reac} \\
    {}_{Z}^{A}X_{N} &\;\rightarrow\; {}_{Z+2}^{A}X_{N-2} + 2e^{-}. \label{eq:0nubb_reac}
    \end{align}

    Of course, the primary focus is on the neutrinoless mode reported in Equation (\ref{eq:0nubb_reac}). To date, it has not been observed, and the most stringent lower bounds on its half-life are at the level of $10^{26}$~yr~\cite{Agostini:2022zub,Gomez-Cadenas:2023}. It can be interpreted as the conversion of two neutrons into two protons and two electrons without the accompanying anti-leptons, effectively corresponding to the creation of matter in the nuclear medium.  This process is directly connected to the established fact, demonstrated by oscillation experiments, that neutrinos possess mass~\cite{Esteban:2020,deSalas:2021,Capozzi:2021}. The SM does not accommodate neutrino masses in its minimal formulation.  To incorporate them, one may supplement the “classical’’ Dirac mass term, analogous to that of the charged fermions, with a right-handed Majorana mass term~\cite{King:2004,Borisov:2024arXiv}. Such a term is allowed by the symmetries of the SM, provided the particle is electrically neutral. Therefore, the existence of a Majorana mass term is not forbidden and well motivated in many SM extensions.  Its inclusion implies that the physical neutrino states are Majorana fermions, indistinguishable from their own antiparticles, which aligns with the theoretical framework proposed by Ettore Majorana nearly a century ago.

\begin{figure}[H]
    \centering
    \includegraphics[width=0.9\textwidth]{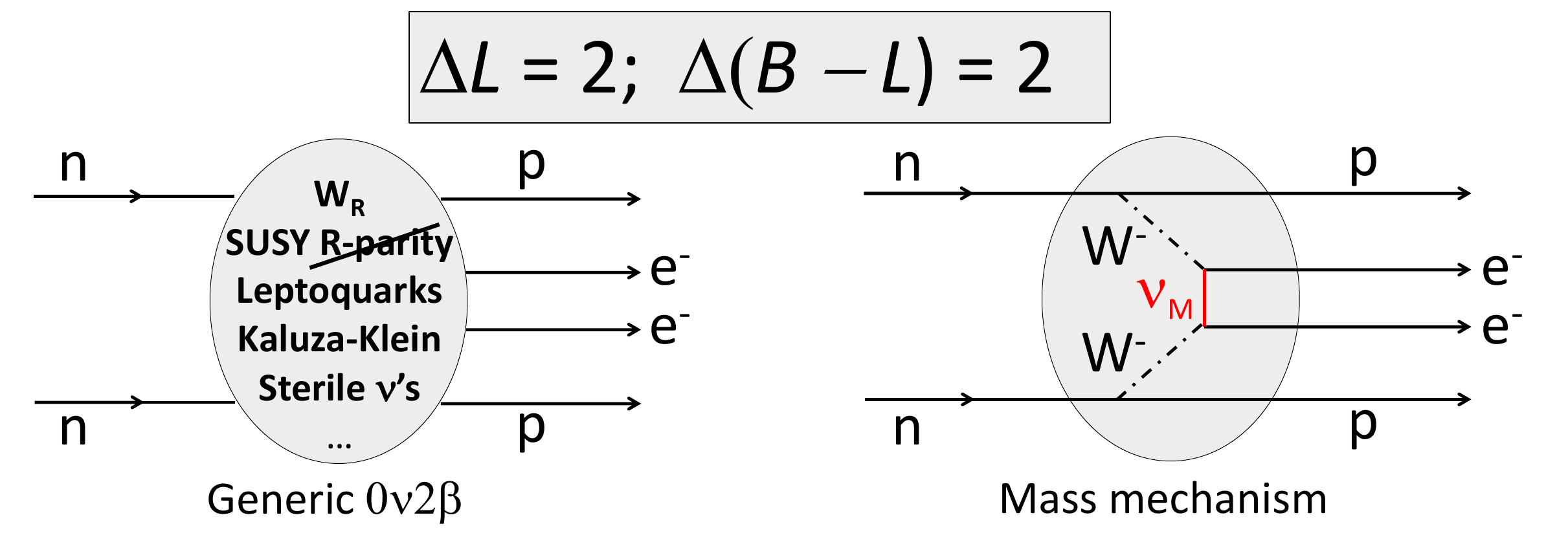}
    \caption{Mechanisms for $0\nu2\beta$. Several possible beyond-Standard-Model channels are schematically shown on the left, while the exchange of a light Majorana neutrino (the mass mechanism) is shown on the right.}
    \label{fig:mechanisms}
    \vspace{-16pt} 
    \end{figure}

    Within this minimal and economical extension of the SM, $0\nu2\beta$ is mediated by the exchange of virtual light Majorana neutrinos: the same particles observed in neutrino oscillation experiments. This scenario is commonly referred to as the ``mass mechanism'' (Figure~\ref{fig:mechanisms}, right panel)~\cite{Bilenky:2015,Vergados:2016}. However, $0\nu2\beta$ may also arise from a broad class of lepton-number-violating interactions beyond the SM~\cite{Pas:2015,Deppisch:2012,Valle:1982}, including the exchange of right-handed gauge bosons, supersymmetric particles with $R$-parity violation, leptoquarks, heavy sterile neutrinos, Kaluza–Klein excitations, and other possibilities discussed in the literature (Figure~\ref{fig:mechanisms}, left panel).  Consequently, beyond its role in neutrino physics, $0\nu2\beta$  constitutes a wide-ranging inclusive probe of lepton number conservation. Even more importantly, its observation would demonstrate the violation of the baryon-minus-lepton number ($B-L$). This quantity represents the only exact, non-anomalous global symmetry of the SM whose violation has not yet been experimentally established. 
\vspace{6pt}
    
    Current interpretations of experimental results are expressed most often within the framework of the mass mechanism, which links $0\nu2\beta$ to fundamental neutrino parameters that are partially accessible through complementary probes~\cite{Bilenky:2015,Vergados:2016}. This framework provides clear experimental benchmarks and enables a meaningful comparison of results obtained with different isotopes, even when experiments rely on distinct methodologies and technologies.  The two central expressions associated with the mass mechanism are the following:
    \begin{align}
    \frac{1}{T_{1/2}^{0\nu}} &= \ln(2)\, G_{0\nu}\, g_A^{4}\, |M_{0\nu}|^{2}\, \left( \frac{m_{\beta\beta}}{m_e} \right)^{\,2},
    \label{eq:rate} \\[6pt]
    m_{\beta\beta} &=
    \bigg|\, m_1\, U_{e1}^{\,2}
    \;+\; m_2\, U_{e2}^{\,2}\, 
    \;+\; m_3\, U_{e3}^{\,2}\, \bigg|,
    \label{eq:mbb}
    \end{align}
    where $T_{1/2}^{0\nu}$ is the $0\nu2\beta$ half-life, $G_{0\nu}$ the exactly calculable phase space, and $|M_{0\nu}|$ the nuclear matrix elements, that sometimes will be denoted NMEs. The axial charge $g_A$ is a parameter of the order of unity, approximately 1.27 for a free neutron. The role of the electron mass $m_e$ is to make $|M_{0\nu}|$ dimensionless, while $G_{0\nu}$ is measured in yr$^{-1}$. The calculation of the NMEs---discussed in Section~\ref{sec:NME}---represents a significant source of uncertainty. 
    
    The parameter $m_{\beta\beta}$ in Equation (\ref{eq:mbb}) is defined as the effective Majorana mass or the effective neutrino mass, and it contains the neutrino mixing matrix elements $U_{ei}$. Importantly, for Majorana neutrinos, the neutrino mixing matrix contains two additional CP-violating phases $\alpha_{21}$ and $\alpha_{31}$ (Majorana phases) that are not accessible in neutrino oscillation experiments, unlike the CP-violating phase $\delta$. They can be made explicit through the expression:
    \begin{equation}
    m_{\beta\beta}
    = \bigg|\, m_1\,c_{12}^2 c_{13}^2
    \;+\; m_2\,s_{12}^2 c_{13}^2\,e^{i\alpha_{21}}
    \;+\; m_3\,s_{13}^2\,e^{i\alpha_{31}} \,\bigg|,
    \label{eq:majorana_phases}
    \end{equation}
    where $c_{ij} \equiv \cos\theta_{ij}$ and  $s_{ij} \equiv \sin\theta_{ij}$ according to the standard parametrization of the neutrino mixing matrix in terms of mixing angles. It should be noted that the values of these phases control the (partial) cancellations among the three contributions in \(m_{\beta\beta}\), so they can suppress or enhance the rate of $0\nu2\beta$ for a given mass spectrum. 
    
    The dependence of the $0\nu2\beta$ rate on the neutrino masses, as expressed by Equations (\ref{eq:rate}) and (\ref{eq:mbb}) can be understood as follows.  In the mass mechanism, a virtual neutrino is emitted at one vertex and reabsorbed at the other, as represented in Figure~\ref{fig:mechanisms}, right panel. Such a process cannot occur if neutrinos are Dirac particles, since the neutrino (absorbed at one vertex) and the antineutrino (emitted at the other one) are distinct in this case. The Majorana nature of neutrinos is therefore essential, but it is not sufficient on its own: the neutrino mass also plays a fundamental role. Because the charged–current interaction has a $V\!-\!A$ structure, the exchanged virtual neutrino is produced and absorbed in its left-chiral component. In the relativistic limit, this leads to a helicity mismatch between emission and absorption. The required opposite-helicity component arises only through a neutrino mass insertion, which suppresses the process and makes the amplitude proportional to the effective Majorana mass $m_{\beta\beta}$. This explains why the decay probability depends on the square of the effective neutrino mass, making $0\nu2\beta$ sensitive to the absolute neutrino mass scale, unlike oscillation experiments. Mixing enters the expression as the electron neutrino involved in the nuclear transition is a superposition of three mass eigenstates, each contributing to the decay amplitude with a weight given by the corresponding mixing matrix element $U_{ei}$.
    
    Another form of neutrinoless process discussed in the literature concerns the emission of a Majoron in the final state~\cite{Gelmini:1981}. Historically, searches for new particles focused on massive and massless bosons called Majorons, the Goldstone bosons that arise from the spontaneous breaking of the global $B-L$ symmetry. In this double beta decay mode, the Majoron would be emitted alongside the two electrons. Experimental searches provide limits on the coupling of neutrinos to the Majoron, but in this review we will not discuss these searches or the corresponding experimental bounds.

    \section{Interplay between Neutrinoless Double Beta Decay and Cosmology}
    \label{sec:consmo}

    The study of $0\nu2\beta$ occupies a central position at the intersection of laboratory neutrino physics and early-Universe cosmology because it probes two ingredients that are essential in many explanations of the cosmic matter–antimatter asymmetry: the violation of total lepton number and the Majorana character of neutrinos. 

    The theoretical framework most commonly invoked to explain the smallness of neutrino masses, which naturally leads to Majorana neutrinos, is the Type-I See-Saw mechanism~\cite{Gell-Mann:1979arXiv,Mohapatra:1980}, implemented by adding a Majorana mass term alongside the Dirac one~\cite{King:2004,Bilenky:2015,Vergados:2016,Borisov:2024arXiv}. On the one hand, this model extends the SM by introducing heavy, sterile, right-handed Majorana neutrinos, denoted $N_R$. On the other hand, the neutrino masses $m_\nu$ of ordinary light neutrinos are generated via the suppression of the electroweak scale by the heavy Majorana mass scale $M_R$, roughly following the relation $m_\nu \sim v^2 / M_R$, where $v$ is the Higgs vacuum expectation value. 
    
    The heavy Majorana neutrinos $N_R$ are unstable and played a crucial role in the thermal history of the early Universe. It is within the decay of these particles that the link to the Baryon Asymmetry of the Universe (BAU) is forged~\cite{Fukugita:1986}. According to the theory of thermal leptogenesis, the heavy $N_R$ states were produced in the hot plasma of the early Universe~\cite{Buchmuller:2005,Davidson:2008}. As the Universe expanded and cooled below the mass threshold of these heavy particles ($T < M_R$), they decayed out of equilibrium into leptons and Higgs bosons~\cite{Sakharov:1991,Buchmuller:2005}. If these decays violate CP symmetry, they can produce a net asymmetry between the number of leptons and antileptons. This CP violation arises from the interference of tree-level and one-loop diagrams involving complex Yukawa couplings~\cite{Covi:1996,Davidson:2008}. Crucially, non-perturbative Standard Model processes known as sphalerons, which are active at temperatures above the electroweak phase transition, violate $B+L$ while conserving $B-L$~\cite{Kuzmin:1985}. These sphalerons can efficiently reprocess the lepton asymmetry generated by the $N_R$ decays into a baryon asymmetry, which explains the observed dominance of matter over antimatter~\cite{Harvey:1990}.

    The connection between $0\nu2\beta$ and this cosmological narrative is centered on the nature of CP violation and the Majorana mass term. A subtle but important issue is the relation between the CP phases $\alpha_{21}$ and $\alpha_{31}$ appearing in Equation~(\ref{eq:majorana_phases}) and those responsible for generating the cosmological asymmetry. In general, these sets of phases are independent. The CP violation driving leptogenesis lies in the Yukawa couplings or in the heavy-neutrino sector. This disconnection is often referred to as the "flavor problem". Only in specially constrained theoretical frameworks, such as models with flavor symmetries or particular See-Saw parameterizations, can a direct relation be established. Thus, observation of $0\nu2\beta$ would demonstrate the existence of Majorana phases but would not automatically determine whether the heavy sector contains the appropriate CP violation required for successful leptogenesis~\cite{Davidson:2008}. 
    
    Another cosmological consideration concerns the timing of lepton-number violating processes in the early Universe. If such processes occur too rapidly while electroweak sphalerons are active, any generated asymmetry can be washed out. The decay observed today occurs at extremely low energies and does not reveal the thermal history of lepton-number violation. Successful leptogenesis requires a delicate interplay between asymmetry generation, CP violation, and washout rates, all of which depend on details of the heavy sector. 

    Heavy Majorana neutrinos with masses $M_I$ could also contribute directly to $0\nu2\beta$ through short-range mechanisms, with amplitudes scaling like $\sum_I V_{eI}^2/M_I$, where $V_{eI}$ are active–sterile mixing elements. Current limits from the decay already restrict certain combinations of heavy masses and mixings, providing valuable information for cosmological scenarios: too-large active–sterile mixing, for example, would lead to excessive washout of any primordial asymmetry~\cite{Deppisch:2012}.

    In summary, a discovery of $0\nu2\beta$ would be transformative: it would prove that neutrinos are Majorana particles and that total lepton number is violated, thereby validating essential ingredients of many leptogenesis models. However, establishing a quantitative link to the cosmological matter–antimatter asymmetry requires additional information about the heavy-neutrino sector or theoretical assumptions that relate low- and high-energy parameters. Until such information is available, the decay remains a necessary but not sufficient laboratory signature for explaining the origin of BAU.

    \section{Factors Governing the Rate of Neutrinoless Double Beta Decay}
    \label{sec:rate}

    The calculations of NMEs and $G_{0\nu}$ are central ingredients in the interpretation of $0\nu2\beta$, since they control the link between the measured half-life and the underlying particle-physics parameters, represented by $m_{\beta\beta}$ in Equation~(\ref{eq:rate}). While $G_{0\nu}$ can be computed with high precision (as discussed in Section~\ref{sec:phase-space}), the calculation of the NMEs presents a significant challenge (Section~\ref{sec:NME}). Their reliable determination is essential for translating experimental limits into constraints on the effective Majorana mass and for comparing the discovery potential of different isotopes. 
    
    \subsection{Calculation of the Nuclear Matrix Elements}
    \label{sec:NME}
    
    The structure of the $0\nu2\beta$ operator, combined with the complexity of correlated nuclear many-body systems, makes the evaluation of NMEs challenging and has motivated a broad range of complementary nuclear-theory approaches~\cite{Vergados:2012,Menendez:2017}. The decay amplitude involves a two-body transition operator connecting the initial and final $0^+$ ground states. For the standard mass mechanism (the only one that will be discussed here), the operator has a characteristic long-range structure, originating from the neutrino propagator, which behaves as $1/r$ at low momentum transfer. In this regime, the transition probes distances of the order of the nuclear size and is particularly sensitive to pairing correlations and the spin–isospin response of the nucleus. The corresponding NME can be decomposed schematically as

    \begin{equation}
    M^{0\nu} = M^{0\nu}_{\rm GT} - \frac{g_V^2}{g_A^2} M^{0\nu}_{\rm F} + M^{0\nu}_{\rm T},
    \label{eq:NME}
    \end{equation}
    where the Gamow–Teller, Fermi, and tensor contributions reflect the underlying weak interaction structure. The Gamow–Teller part dominates, but all terms must be treated consistently, including higher-order nucleonic currents and finite nucleon size effects. In contrast, mechanisms involving heavy mediators lead to short-range operators suppressed by the mediator mass, and therefore require an accurate treatment of short-distance physics, nuclear contact operators, and nucleon–nucleon correlations beyond simple phenomenological prescriptions~\cite{Cirigliano:2018}.

    Any $0\nu2\beta$ calculation must start from a reliable nuclear Hamiltonian and a model for the initial and final nuclear states. The decay operator involves two nucleons converting into two protons with the emission of two electrons (and no neutrinos), mediated by the exchange of a virtual Majorana neutrino. The amplitude therefore depends not only on single-particle transition matrix elements but also on two-body correlations in the nuclear medium. In all approaches one has to account for~\cite{Vergados:2012,Vergados:2016,Menendez:2017}:  
\begin{itemize}[leftmargin=*, labelsep=5.8mm, itemsep=0pt, topsep=3pt, partopsep=0pt, parsep=0pt]
      \item The proper treatment of short-range correlations between nucleons.  
      \item The appropriate radial dependence of the neutrino propagator (often approximated by a closure approximation). 
      \item Corrections from finite nucleon size and higher-order terms in the nucleon currents (e.g. induced pseudoscalar, weak magnetism). 
      \item The proper normalization of nuclear wave functions for initial and final even–even nuclei.  
    \end{itemize}

    Because $0\nu2\beta$ involves a large momentum transfer (the order of a few hundred MeV), the neutrino potential is very sensitive to these details, which makes the NME calculations significantly more demanding than those for $2\nu2\beta$. Four main frameworks have been developed so far to compute $0\nu2\beta$ NMEs~\cite{Menendez:2017}:  
\begin{itemize}[leftmargin=*, labelsep=5.8mm, itemsep=0pt, topsep=3pt, partopsep=0pt, parsep=0pt]
      \item {{Nuclear Shell Model (NSM)}
}~\cite{Menendez:2009}. Sometimes denoted also as Interacting Shell Model (ISM), this method picks a limited set of active shells (single-particle orbitals) for protons and neutrons, and diagonalizes the full nuclear Hamiltonian in this truncated space. The NSM can treat complex correlations among valence nucleons very accurately, including pairing, configuration mixing, and deformation within the model space. However, since the model space is often limited, contributions from outside the valence shells (“core” or “higher-shell excitations”) must be approximated or neglected, which can lead to underestimation of the NME.  
      \item {{Quasiparticle Random Phase Approximation (QRPA)}}~\cite{Simkovic:2008}. This method starts from a BCS-like description of pairing for protons and neutrons, builds quasiparticle excitations, and treats the two-body decay operator in a linearized (RPA) framework. QRPA can include many single-particle orbitals, hence better incorporate high-energy excitations and contributions from many shells; long-range correlations are easier to include. The drawback is that QRPA often relies on adjustable parameters (e.g. coupling strengths, particle–particle interaction strength), and results can vary sensitively with their choice.  
      \item {{Generating Coordinate Method (GCM)}}~\cite{Rodriguez:2010}. Also known as the generator-coordinate method, this approach is implemented on top of self-consistent mean-field solutions provided by an energy-density functional (EDF). The EDF supplies a set of intrinsic Hartree–Fock–Bogoliubov configurations characterized by collective coordinates such as quadrupole deformation or pairing strength. GCM builds correlated nuclear states by restoring symmetries (particle number, angular momentum) and mixing these configurations with variationally determined weights. This EDF-GCM framework captures large-amplitude collective motion (including shape coexistence and deformation mixing) and incorporates correlations beyond those accessible in simple mean-field, shell-model valence spaces, or QRPA formulations.
      \item {{Interacting Boson Model (IBM)}}~\cite{Iachello:2013}. This method maps pairs of nucleons (proton-proton, neutron-neutron, proton-neutron) to bosons, and describes the collective nuclear dynamics in terms of boson wave functions. IBM is particularly useful for medium to heavy nuclei where full shell-model diagonalization is impractical. Its phenomenological nature can, however, introduce systematic uncertainties depending on the boson mapping and the model parameters.  
    \end{itemize}

    \textls[-10]{Because each method treats correlations and nuclear dynamics differently, the resulting NMEs can differ significantly for the same isotope~\cite{Menendez:2017}. These differences lead to sizable theoretical uncertainties in the value of the NMEs, which remain the main limiting factor in the particle-physics interpretation of $0\nu2\beta$ experiments. Concretely:}

\begin{itemize}[leftmargin=*, labelsep=5.8mm, itemsep=0pt, topsep=3pt, partopsep=0pt, parsep=0pt]
      \item For a given isotope, different methods can yield values of $|M_{0\nu}|$ that differ by up to a factor of two (or more), implying up to an order-of-magnitude uncertainty in the inferred effective Majorana mass $m_{\beta\beta}$ from a measured (or lower-bound) half-life.  
      \item Part of this uncertainty stems from treatment of short-range correlations and nuclear current corrections (finite nucleon size, higher-order terms), which are handled differently in different frameworks.  
      \item \textls[-15]{Another source is the limited model space (in NSM) or approximations in correlation strength (in QRPA or IBM).  }
    \end{itemize}

    To reduce uncertainties, ongoing efforts include: improving nuclear Hamiltonians, expanding model spaces, including more realistic two-body currents and nucleon-nucleon correlations, and benchmarking among models. It is also important to use realistic single-particle energies, consistent pairing interactions, and self-consistent treatment of deformation and configuration mixing across initial and final nuclei.  

    An important recent development is the explicit separation of long- and short-range contributions to the NME~\cite{Cirigliano:2018}. The long-range part is controlled by low-momentum nuclear structure, where conventional models are reliable, while the short-range sensitivity requires a consistent treatment of nucleon–nucleon repulsion, two-body currents, and contact operators derived from effective field theory. Different many-body approaches incorporate these ingredients in varying ways, and coordinated efforts are under way to benchmark and harmonize their treatments. Effective field theory provides a systematic and model-independent framework, enabling a consistent separation of long- and short-range physics and an ordered inclusion of many-body currents and contact operators~\cite{Brase:2022}.

    Significant progress has been achieved in \emph{ab initio} nuclear theory~\cite{Menendez:2017,Yao:2020,Belley:2021}, which aims to compute nuclear observables starting from inter-nucleon interactions constrained by chiral effective field theory. For $0\nu2\beta$ decay, \emph{ab initio} techniques have already provided matrix elements for light systems and for Gamow–Teller and double-beta operators in nuclei up to the $sd$ and lower $pf$ shells. These results illuminate the role of two-body weak currents, short-range correlations, and operator renormalization. Although full $0\nu2\beta$ NMEs for the heaviest candidate isotopes remain beyond current computational reach, the \emph{ab initio} benchmark calculations guide the construction and validation of effective operators in traditional many-body spaces, and are beginning to reduce long-standing ambiguities in quenching and operator renormalization.
    
    For heavier isotopes of experimental interest such as $^{76}$Ge, $^{100}$Mo, $^{130}$Te or $^{136}$Xe, state-of-the-art many-body approaches continue to refine their predictions~\cite{Menendez:2017}. Large-scale QRPA and IBM calculations incorporate improved treatment of isospin restoration and particle–particle interactions. Shell-model studies now employ expanded valence spaces and non-perturbative operator renormalization informed by \emph{ab initio} results. EDF-based calculations are including beyond-mean-field correlations through symmetry restoration and configuration mixing. Cross-comparisons of these methods, together with systematic uncertainty estimates, constitute a major advancement relative to the situation a decade ago.
    
    In Equation~(\ref{eq:rate}), an important role is played by the axial-vector coupling constant $g_A$. It is well known that many nuclear-structure calculations of ordinary $\beta$ decay and $2\nu2\beta$ (based for example on QRPA methods) require a quenched value of $g_A$ to reproduce experimental data, typically reducing $g_A$ from its free-nucleon value of 1.27 to an effective value in the range $0.7$--$1.0$~\cite{Menendez:2017,Suhonen:2020}. This reduction would lead to a remarkable increase of the predicted half-lives, decreasing $0\nu2\beta$ sensitivity to the effective Majorana mass. This quenching is often attributed to missing correlations in truncated model spaces and to two-body currents. However, QRPA calculations of ordinary muon capture (OMC) at momentum transfers of the order of $100~\text{MeV}$ have been shown to reproduce the experimental capture rate without requiring any quenching of $g_A$~\cite{Simkovic:2020}. This finding is further supported by recent measurements of OMC on $^{76}$Se in the MONUMENT experiment~\cite{Araujo:2025arXiv}. Since $0\nu2\beta$ via light-Majorana-neutrino exchange also involves virtual momentum transfers of similar magnitude, this result provides theoretical support for employing the free-nucleon value of $g_A$ in $0\nu2\beta$ nuclear matrix element calculations. Although encouraging, this conclusion is not universal, because OMC and $0\nu2\beta$ probe different operator structures, nuclear correlations, and multipole compositions, and different nuclear models may not exhibit the same momentum-transfer dependence. In this article however, following the convention generally adopted in literature and by the double-beta decay community, we will use the free-nucleon value for $g_A$. We also note that \emph{ab initio} calculations, by their very construction, do not require any $g_A$ quenching.

    Because the half-life for $0\nu2\beta$ scales roughly as $|M_{0\nu}|^2$ (Equation~(\ref{eq:rate})), the uncertainty in \(M_{0\nu}\) propagates quadratically into the extraction of the effective Majorana mass $m_{\beta\beta}$~\cite{Menendez:2017}. This means that, even with a very high-statistics and low-background experiment, our ability to infer neutrino mass parameters will remain limited unless nuclear theory advances reduce the present spread in NME values. Therefore, nuclear-theory uncertainty is the dominant systematic in translating experimental results into constraints on fundamental neutrino properties. Reducing this uncertainty remains a key goal for the community if future experimental efforts are to deliver definitive insight into the absolute neutrino mass scale. 
    
\begin{figure}[H]
    \centering
    \includegraphics[width=1\textwidth]{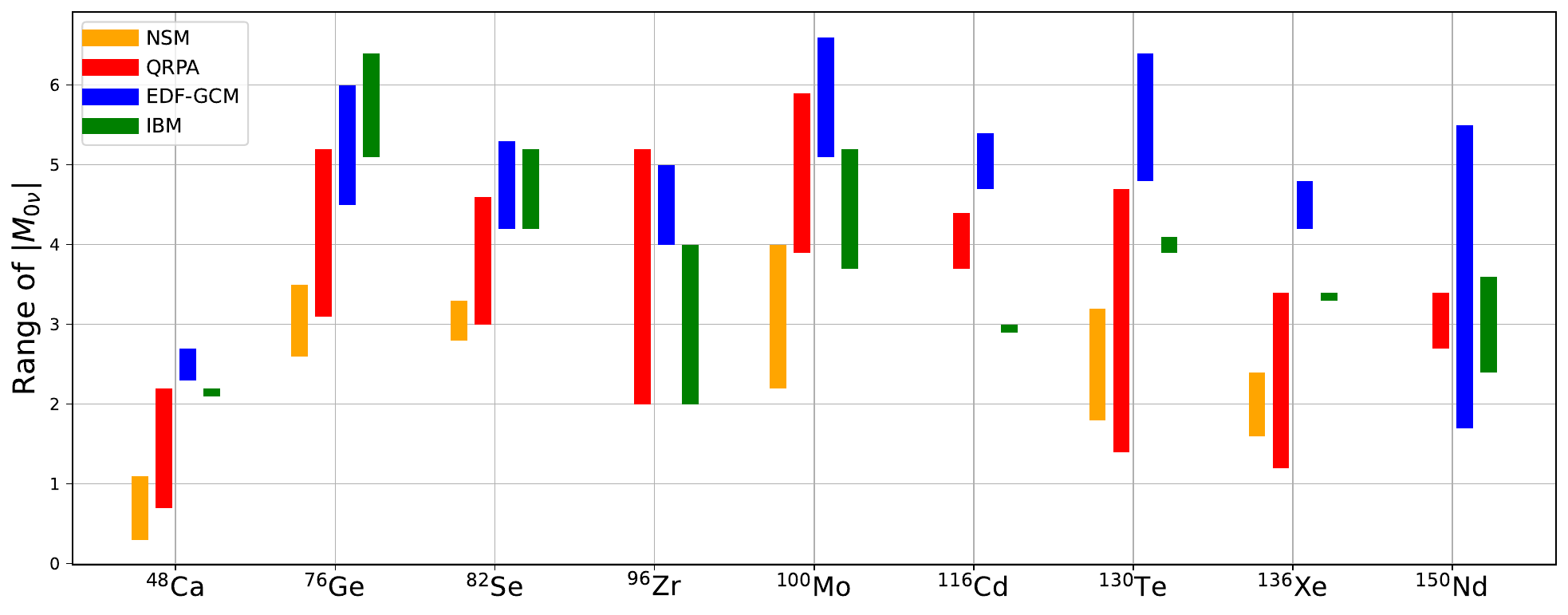}
    \caption{Ranges of $|M_{0\nu}|$ values for the ``magnificent nine" experimentally relevant $0\nu2\beta$ candidates as predicted by the four main nuclear-theory frameworks used for their computation.}
    \vspace{-12pt}
    \label{fig:NME-spread}
    \end{figure}

    The spread in calculation can be appreciated quantitatively in Figure~\ref{fig:NME-spread}, which reports representative values drawn from major reviews and widely used reference calculations for a set of experimentally relevant double-beta decay candidates~\cite{Agostini:2022zub,Menendez:2017,Menendez:2009,Simkovic:2008,Rodriguez:2010,Iachello:2013,Coraggio:2022}. (I will refer to these nuclei as the “magnificent nine,” as they stand out among the 22 $\beta^-\beta^-$ candidates as the most favorable for a successful experimental search, as will be discussed in more detail in Section~\ref{sec:isotope}.) Shell-model (NSM) results are systematically lower due to restricted valence spaces, whereas QRPA spans a broader interval influenced by choices of particle--particle strength, short-range correlations, and isospin restoration. EDF-GCM calculations generally yield larger NMEs. IBM values typically lie between NSM and EDF-GCM, and often track the EDF-GCM trend. Differences across methods reflect model-space truncations, operator choices, deformation effects, and the treatment of correlations. The features of some isotopes push the shell-model computational problem beyond its present limits, hence the NSM entry is omitted. It is important to note that these values serve only as indications of the spread and general trends; they should not be used to extract precise values of effective Majorana masses from half-life measurements. For accurate determinations, one must refer to the specific literature pertaining to each isotope or technical reviews.

\vspace{-16pt}
    \subsection{Calculation of the Phase Space}
    \label{sec:phase-space}

   The phase--space factor $G_{0\nu}$ in Equation~(\ref{eq:rate}) collects all purely leptonic kinematic effects, including the distortion of the outgoing electrons by the Coulomb field of the daughter nucleus.  It depends on the decay $Q$-value and on the charge $Z+2$ of the final nucleus.  The $Q$-value is defined as
    \begin{equation}
    Q \equiv M({}_{Z}^{A}X_{N}) - M({}_{Z+2}^{A}X_{N-2}) ,
    \label{eq:Q-value}
    \end{equation}
    with $M({}_{Z}^{A}X_{N})$ and $M({}_{Z+2}^{A}X_{N-2})$ the atomic masses of the parent and daughter.  This definition ensures that $Q$ is the total kinetic energy available to the two emitted electrons in the $0\nu2\beta$ channel.

    A general and widely used expression for the phase--space factor (PSF) in the standard light-Majorana mechanism is a double integral over the electron energies,
    \begin{equation}
    G_{0\nu}(Q,Z+2)
      = \frac{G_F^4\cos^4\theta_C}{(2\pi)^5}
        \!\int_{m_e}^{Q-m_e}\! dE_1
        \!\int_{m_e}^{Q-E_1}\! dE_2 \;
           p_1 E_1 \, p_2 E_2 \,
           F(Z+2,E_1)\,F(Z+2,E_2)\,
           \mathcal{K}(E_1,E_2),
    \label{eq:PSF}
    \end{equation}
    where $E_i$ and $p_i=\sqrt{E_i^2-m_e^2}$ are the energies and momenta of the two electrons, $F(Z+2,E)$ describes the Coulomb distortion of the electron wave functions, and $\mathcal{K}(E_1,E_2)$ is the leptonic kinematic kernel which depends on the particular $0\nu$ operator responsible for the decay.  Different normalization appear in the literature, but the integral structure and physical content remain consistent.

    The reliability of $G_{0\nu}$ depends on the accuracy of the electron wave functions. Early plane--wave or simple Fermi--function descriptions capture only crude Coulomb effects. Modern calculations~\cite{Kotila:2012} instead solve the Dirac equation in the finite--size Coulomb field of the daughter nucleus and include atomic screening; this provides accurate large/small components of the electron spinors in the nuclear volume and is now the standard approach. Finite--size and screening effects are essential, with additional small refinements (such as radiative corrections) included when needed. Nonstandard mechanisms modify only the kernel $\mathcal{K}(E_1,E_2)$, while the overall PSF framework remains unchanged. $\mathcal{K}(E_1,E_2) = 1$ for the mass mechanism up to small recoil or higher order corrections.

    Practically, one specifies the isotope, constructs the screened finite--size Coulomb potential, solves the Dirac equation for continuum electrons at all energies entering the integration, evaluates the radial wave functions at (or over) the nuclear volume, inserts them into the integrand together with the appropriate $\mathcal{K}(E_1,E_2)$, and finally performs the double integration with careful treatment of the kinematic endpoints. Because conventions differ among authors, quoted values of $G_{0\nu}$ must always state the adopted normalization and inputs. Modern PSF evaluations are considered robust and reliable and form a solid component of present analyses of $0\nu2\beta$ searches.
    
\begin{figure}[H]
    \centering
    \includegraphics[width=0.8\textwidth]{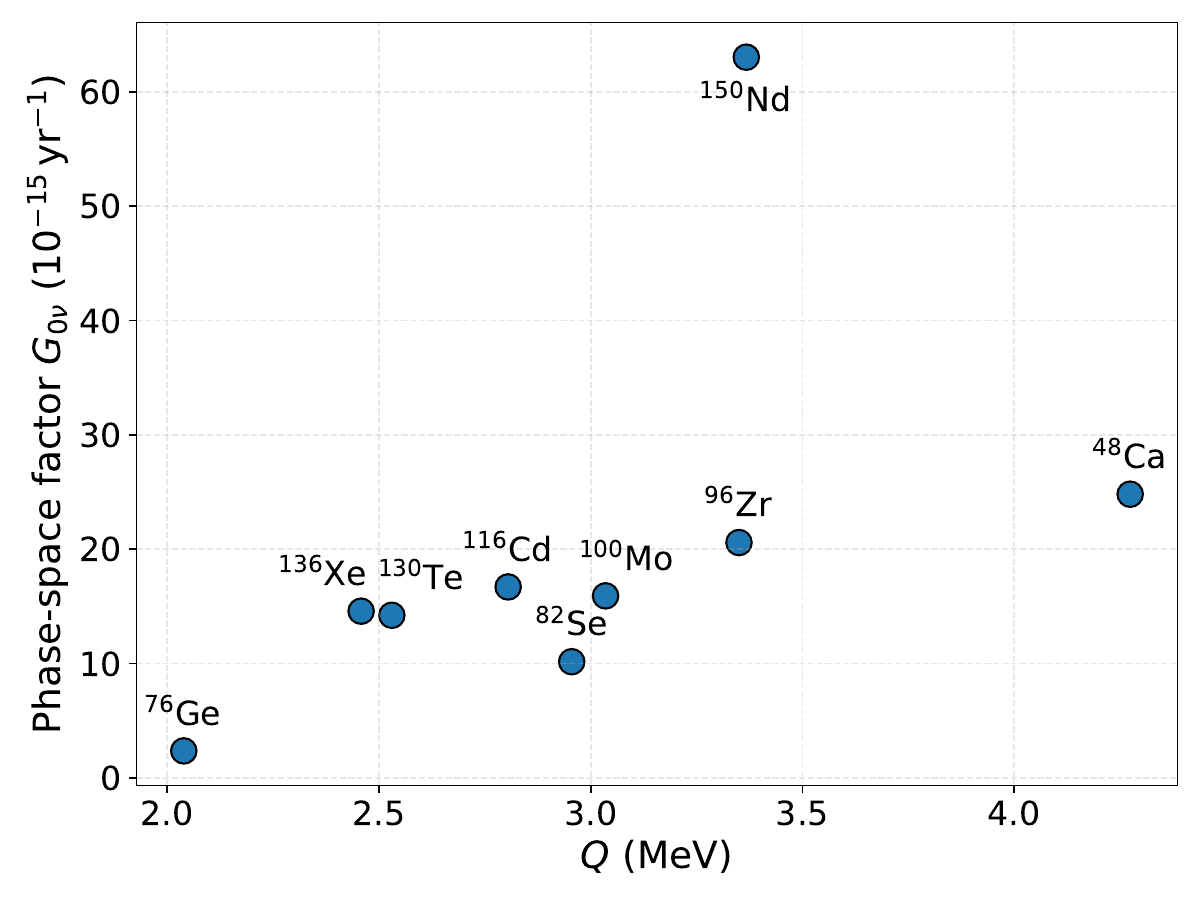}
    \caption{Phase-space factors for the ``magnificent nine" experimentally relevant $0\nu2\beta$ candidates as a function of the $Q$-value. Values are extracted from {Ref.}
~\cite{Kotila:2012}.}
    \label{fig:PSFvsQ}
    \vspace{-20pt} 
    \end{figure}

    The dependence of the PSF on the released energy is very strong: in a polynomial approximation, the leading term behaves as $G_{0\nu} \sim Q^{5}$, reflecting the rapid expansion of the PSF with increasing $Q$-value. As a consequence, isotopes with higher $Q$ benefit from substantially larger PSFs, which enhances their intrinsic sensitivity to $0\nu2\beta$. The PSF as a function of the $Q$-value is shown for the “magnificent nine’’ isotopes in Figure~\ref{fig:PSFvsQ}~\cite{Kotila:2012}. Most candidates exhibit $G_{0\nu}$ values in the range $(10$–$20)\times 10^{-15}\,\text{yr}^{-1}$, with two notable exceptions: $^{76}\mathrm{Ge}$, whose PSF is only $\sim\!2.4$ in these units (due to its comparatively low $Q$-value), and, on the opposite end, $^{150}\mathrm{Nd}$, which reaches $\sim\!63$. A useful way to understand why $^{150}\mathrm{Nd}$ has an unusually large PSF is to recall that, in simple analytic estimates, the leading contribution scales approximately as $G_{0\nu} \propto Z^{2} Q^{5}$. While its high $Q$-value already provides a strong enhancement, $^{150}\mathrm{Nd}$ also has one of the largest nuclear charges among the relevant candidates.

    \section{Experimental Strategies}
    \label{sec:strategies}

    Once the signature of $0\nu2\beta$ has been presented and discussed, we will report on the choice of candidate nucleus and detection technique, two strongly interdependent aspects. This section reviews the approaches that have led to the design of today’s most sensitive experiments.

     \subsection{Experimental Signatures and the Role of Two-Neutrino Double Beta Decay}
     \label{sec:signatures}
     
     As $0\nu2\beta$ is characterized by the emission of two electrons from a nucleus without the emission of neutrinos, the minimal required signature for this process is the detection of the two electrons and the measurement of their sum energy $E_{\rm sum}$. Since no neutrinos carry away energy in the $0\nu$ mode, the sum energy of the two electrons is, to an excellent approximation, equal to the transition $Q$-value. In an ideal detector with perfect resolution, the observable signature of $0\nu2\beta$ is therefore a monoenergetic peak at $Q$. 
    
\begin{figure}[H]
            \centering
            \includegraphics[width=0.7\linewidth]{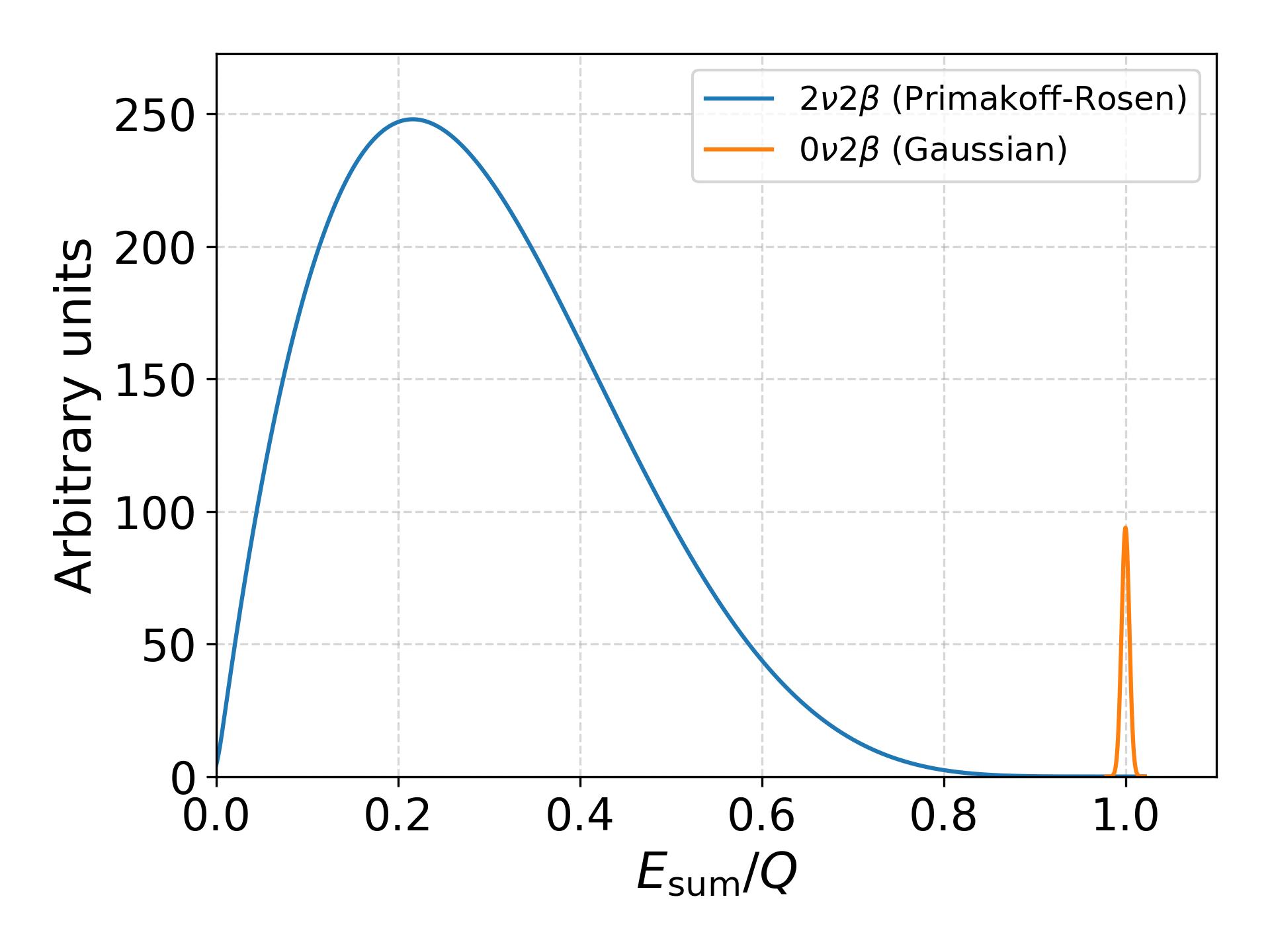}
\vspace{-6pt}
            \caption{The summed--energy spectrum of the two electrons from the $2\nu2\beta$, described within the Primakoff--Rosen approximation, is shown together with the monoenergetic peak expected from $0\nu2\beta$. A Gaussian detector response with a FWHM energy resolution of 1\% is assumed. The ratio between the $2\nu2\beta$ and $0\nu2\beta$ decay rates is unrealistically set as low as 100 to highlight the $0\nu2\beta$ peak.}
            \label{fig:0n-vs-2n}
    \end{figure}

    In contrast, in the standard $2\nu2\beta$ the neutrinos carry away a variable amount of energy, producing a continuous distribution of $E_{\rm sum}$ between 0 and $Q$. A commonly used analytic expression for the summed–electron energy spectrum of $2\nu2\beta$ is obtained by the simplified but realistic Primakoff-Rosen approximation, which includes leading Coulomb effects on the outgoing electrons in a simple way~\cite{Primakoff:1959}. In this approximation the spectral shape can be written as:

    \begin{equation}
    \frac{d\Gamma_{2\nu}}{dE_{\rm sum}}
    =
    \mathcal{N}\,
    E_{\rm sum}
    \left(E_{\rm sum}^4 + 10 E_{\rm sum}^3 + 40 E_{\rm sum}^2 + 60 E_{\rm sum} + 30 \right)
    (Q - E_{\rm sum})^5,
    \label{eq:Primakoff-Rosen}
    \end{equation}
    where the constant $\mathcal{N}$ absorbs numerical factors, phase-space constants, NMEs and the weak interaction coupling $G_F^4 \cos^4\theta_C$. From Equation~(\ref{eq:Primakoff-Rosen}), one finds that $S_{2\nu}(E)$ vanishes at the endpoint according to a power law. Near $E \rightarrow Q$, the spectrum behaves as $(Q -E_{\rm sum})^n$ with $n$ typically around $5$. The qualitative difference is therefore a delta-like peak at $Q$ for $0\nu2\beta$ and a broad continuum extending up to $Q$ for $2\nu2\beta$, illustrated in Figure~\ref{fig:0n-vs-2n}. 
\vspace{-6pt}

\vspace{-16pt}
  
\begin{figure}[H]
            \centering
            \includegraphics[width=0.7\linewidth]{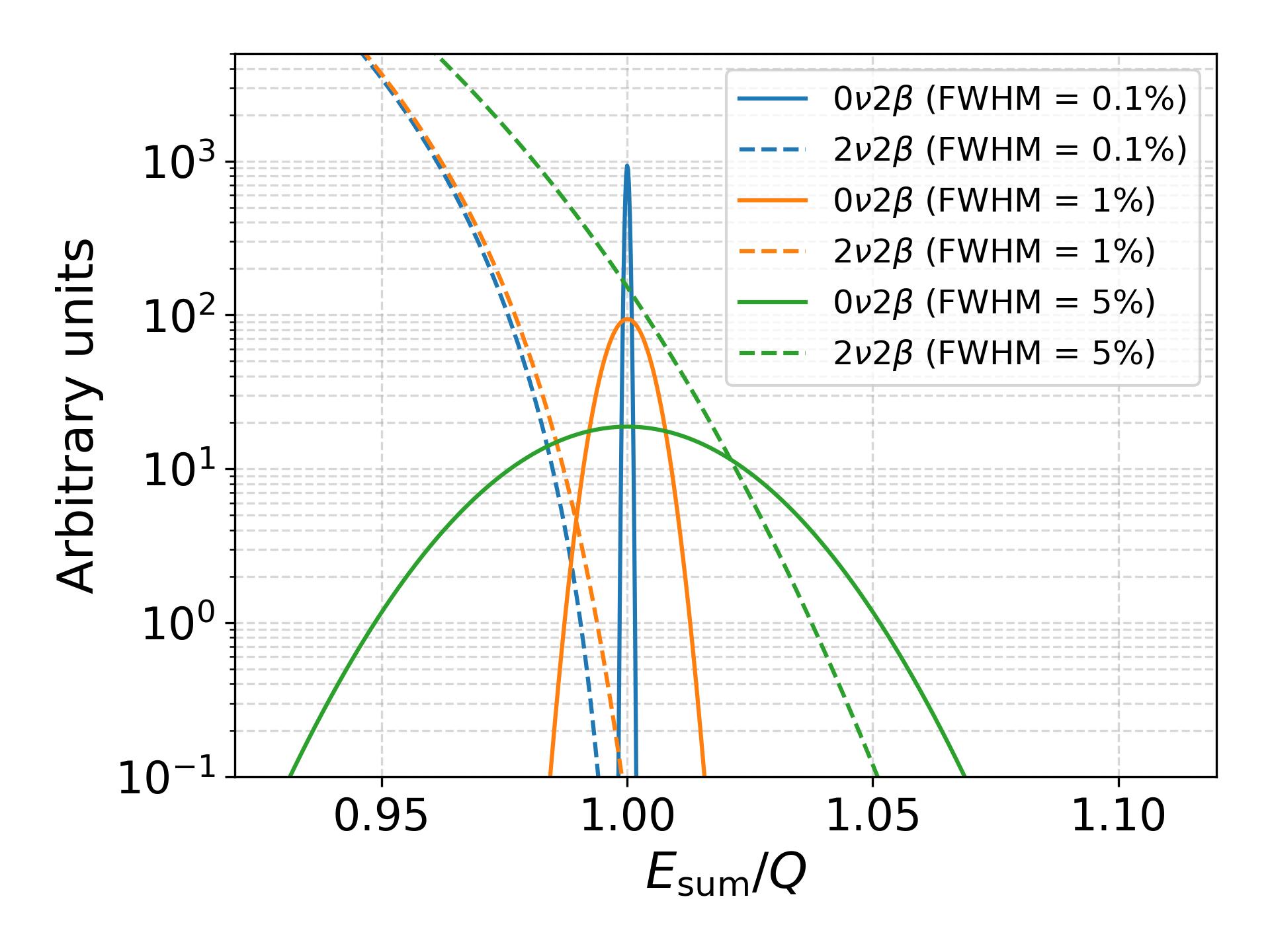}
\vspace{-6pt}
            \caption{For three different energy resolutions (assuming a Gaussian detector response), the high--energy tail of the $2\nu2\beta$ spectrum, convolved with the energy resolution, is shown together with the corresponding $0\nu2\beta$ peak broadened by the same resolution. The ratio between the $2\nu2\beta$ and $0\nu2\beta$ decay rates is set to $10^{8}$, consistent with the sensitivities expected for next--generation experiments.}
            \label{fig:0n-vs-2n-3sigmas}
    \end{figure}

    In a real detector, the observed spectrum is the convolution of the true spectrum with the detector response function. With a Gaussian energy response, the $0\nu2\beta$ peak becomes a Gaussian centered at $Q$ (Figure~\ref{fig:0n-vs-2n}), while the continuum of the $2\nu$ spectrum gets smeared, acquiring a small but nonzero tail at energies close to $Q$. This implies that $2\nu2\beta$ can become a background for $0\nu2\beta$ if the detector energy resolution is insufficient. The amount of $2\nu2\beta$ leakage into a narrow region of interest (ROI) around $Q$ depends strongly on the energy resolution. Approximating the endpoint spectrum as $(Q-E_{\rm sum})^n$ and using a Gaussian response, the number of leaked events scales as $(\Delta E_{FWHM})^{n+1}$ where $\Delta E_{FWHM}$ is the full-width-half-maximum of the Gaussian peak. For the typical case $n \approx 5$, this gives $(\Delta E_{FWHM})^{6}$, illustrating how crucial excellent energy resolution is for suppressing this background. This is shown in Figure~\ref{fig:0n-vs-2n-3sigmas}, where we consider a realistic $0\nu2\beta$ decay rate, assumed to be $10^{-8}$ times smaller than the typically measured $2\nu2\beta$ rate (corresponding to the half-life sensitivities targeted by next-generation experiments, which are in the range $10^{27}$-$10^{28}$~yr). One observes that, for the energy resolutions achievable with germanium diodes and thermal bolometers ($\Delta E_{FWHM} = 0.1\%$), the background contribution from $2\nu2\beta$ decay is negligible. While an energy resolution degraded by one order of magnitude remains acceptable, a resolution of $5\%$ results in a ROI that is fully dominated by the high--energy tail of the $2\nu2\beta$ spectrum.
\vspace{-6pt}

\vspace{-16pt}

    A second way in which $2\nu2\beta$ can contribute to the background arises from poor time resolution~\cite{Chernyak:2012}. If two independent $2\nu$ decays occur within a time interval $\tau$ that the detector cannot resolve, their pulses overlap and may be reconstructed as a single event with energy approximately equal to the sum of the two individual energies. The accidental coincidence rate is proportional to $r^2 \tau$, where $r$ is the single-event rate. Only a small fraction of these pile-up events will fall into the $0\nu$ ROI, but their contribution can be non-negligible in detectors with slow response (e.g. thermal bolometers) unless specific pulse-shape or multiplicity-rejection techniques are used.
    
    In summary, the unmistakable signature of $0\nu2\beta$ is the detection of two electrons with a sum energy equal to the $Q$-value. The irreducible background from the allowed two-neutrino mode is suppressed by energy resolution, which limits the spectral leakage near $Q$, and by time resolution, which suppresses accidental pile-up. For the most important isotopes, the $Q$-value is known a priori with sub-keV precision from high-resolution Penning-trap mass spectrometry, so the energy of the decay peak to be searched for is precisely determined in advance. 
    
    In addition to this primary signature, further information can be obtained from the single-electron energy spectrum, the angular correlation between the two emitted electrons, and the identification of the recoiling nucleus' atomic species. 

    \subsection{Which Isotopes are Best Suited for Neutrinoless Double Beta Decay?}
    \label{sec:isotope}

    Several decades of experimental activity have shown that the choice of isotope is a primary design driver for any $0\nu2\beta$ experiment. In practice, three criteria dominate this choice:
\begin{enumerate}[leftmargin=2.3em, labelsep=4mm, label=(\roman*), itemsep=0pt, topsep=3pt, parsep=0pt]
    \item the $Q$-value of the transition,
    \item the natural isotopic abundance and the feasibility of enrichment,
    \item the compatibility with a high-performance detection technique.
    \end{enumerate}

    The $Q$-value is arguably the most important parameter, as it simultaneously governs the available phase space and the exposure to natural radioactive backgrounds. The role of the $Q$-value in terms of phase-space factor $G_{0\nu}$ was extensively discussed in Section~\ref{sec:phase-space} and is illustrated in Figure~\ref{fig:PSFvsQ} for the nine most favorable isotopes, introduced above as the ``magnificent nine''. Their $Q$-values exceed 2.4~MeV, with the notable exception of ${}^{76}$Ge ($Q=2039$~keV), which remains a prime candidate owing to the outstanding performance achievable with germanium detectors.

\begin{figure}[H]
    \centering
    \vspace{-5pt}
    \includegraphics[width=0.8\textwidth]{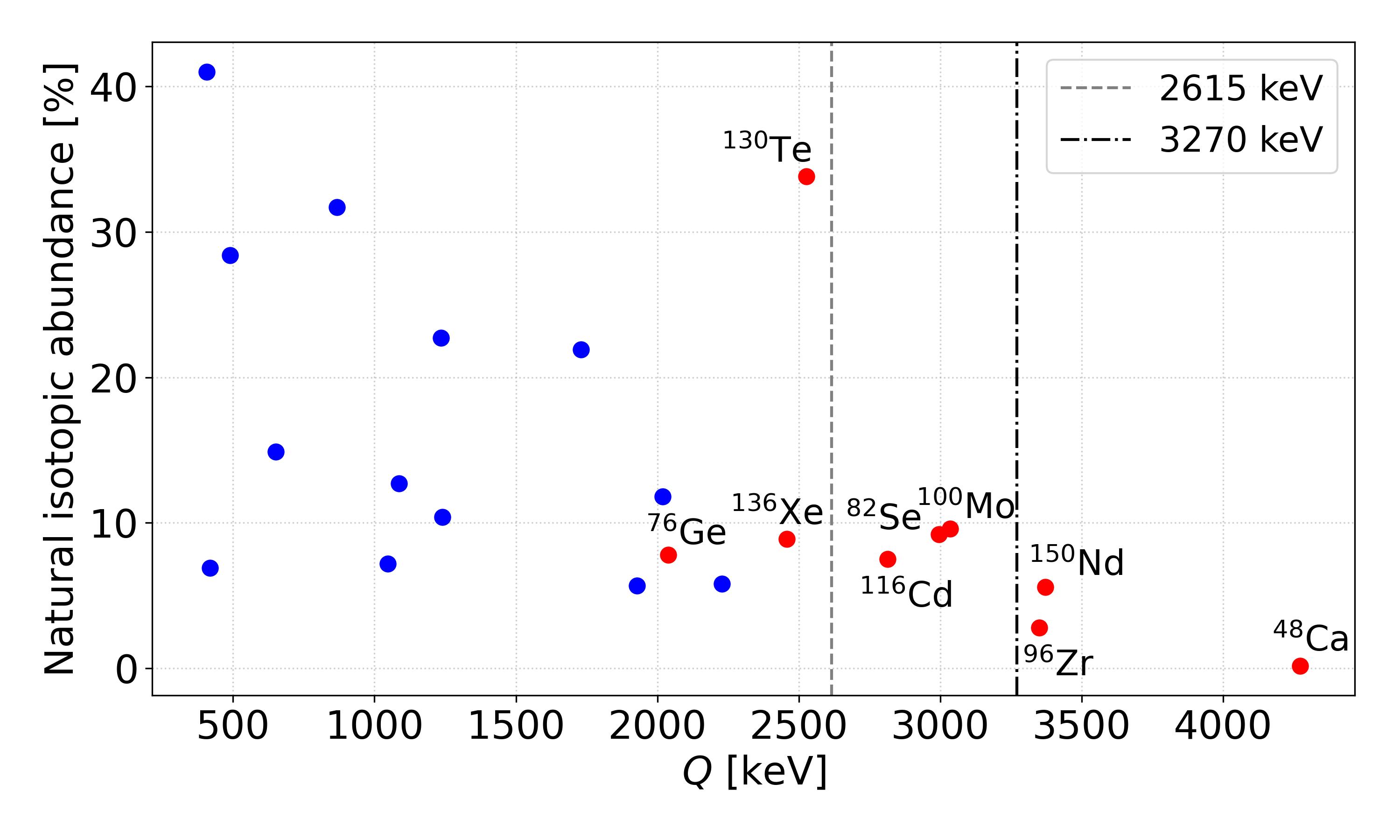}
\vspace{-6pt}
    \caption{Natural isotopic abundance and $Q$-value for the 22 $0\nu2\beta$ candidates. The ``magnificent nine'' experimentally relevant isotopes are specified and highlighted in red. The two vertical markers show energy thresholds connected to $\gamma$ radioactivity (2615~keV) and radon-related $^{214}$Bi $\beta$ radioactivity (3270~keV).}
    \label{fig:candidates}
    \end{figure}

    Figure~\ref{fig:candidates} summarizes the situation for all 22 known $\beta^-\beta^-$ nuclei, showing their natural isotopic abundances versus the corresponding double-beta decay $Q$-values. The ``magnificent nine'' are highlighted in red, and two vertical markers indicate important background-related thresholds: the 2615~keV $\gamma$-ray line from natural radioactivity and the 3270~keV endpoint of the ${}^{214}$Bi $\beta$ decay, the most energetic among the ${}^{222}$Rn daughters. These markers naturally divide the ``magnificent nine'' into three groups. The first group (${}^{76}$Ge, ${}^{130}$Te, ${}^{136}$Xe) lies below 2615~keV and must contend with residual environmental $\gamma$-ray background as well as radon-induced activity. The second group (${}^{82}$Se, ${}^{100}$Mo, ${}^{116}$Cd) is largely beyond the bulk of natural $\gamma$ background, although radon progeny may still contribute. The third group (${}^{48}$Ca, ${}^{96}$Zr, ${}^{150}$Nd) lies above the ${}^{214}$Bi endpoint and is, in principle, in the most favorable position for background-free experiments.

\vspace{-16pt}

    Figure~\ref{fig:candidates} also shows natural isotopic abundances. Most candidate isotopes occur at the level of a few percent, with two notable exceptions: the favorable case of ${}^{130}$Te, whose abundance of 34.1\% allows high-sensitivity searches even with natural material, and the unfavorable case of ${}^{48}$Ca, whose abundance is well below 1\%. As discussed below, frontier experiments aiming to explore half-lives beyond $10^{27}$--$10^{28}$~yr typically require isotope masses of at least $\mathcal{O}(100~\mathrm{kg})$, often approaching and surpassing the ton scale depending on the isotope and the technique. Since detector size and background scale roughly with the total source mass rather than the isotope mass alone, isotopic enrichment is indispensable for nearly all high-sensitivity searches.
   
    Among the available enrichment technologies, gas centrifugation is the only method that combines industrial maturity, large production capacity, and acceptable cost. Its applicability is limited to elements that form suitable gaseous compounds, usually fluorides. This restricts its use to ${}^{76}$Ge, ${}^{82}$Se, ${}^{100}$Mo, ${}^{116}$Cd, and ${}^{130}$Te, while ${}^{136}$Xe is naturally gaseous. 
    Enrichment costs, which depend strongly on the supplier, on the amount purchased, and on commercial negotiations, typically carry an uncertainty up to a factor of 2; for reference, germanium, selenium, and molybdenum cost roughly 50--100 \$/g, while cadmium is about 3$\times$ more expensive, and tellurium ($\sim$\!0.2$\times$) and xenon ($\sim$\!0.1$\times$) are less expensive, with similar uncertainty.
    Xenon is by far the most economical and scalable option, benefiting from large commercial demand, high production capacity, and the possibility of isotope recovery and reuse. The procurement of enriched double-beta decay isotopes from Russia---long a dominant producer due to its mastery of stable isotope enrichment---faces obvious geopolitical difficulties; however, substantial production exists in the West, such as large-scale enrichment of selenium and germanium by URENCO, and a major stable-isotope enrichment facility was recently established by ORANO in France. In addition, significant productions of enriched molybdenum and enriched germanium were achieved in China by IPCE and CNEIC, respectively (both subsidiaries of CNNC). For a recurring conspiracy of nature, the three isotopes with the most favorable $Q$-values---${}^{48}$Ca, ${}^{96}$Zr, and ${}^{150}$Nd---cannot currently be enriched industrially by centrifugation. Alternative techniques such as electromagnetic separation, ion cyclotron resonance~\cite{Dolgolenko:2009}, and laser-based methods such as AVLIS~\cite{Stern:1985} have demonstrated technical feasibility but remain at the R\&D or pilot-production level. Recently, zirconium enrichment by gas centrifugation was obtained, but throughput and cost remain uncertain.
    
    The importance of the third criterion---the compatibility between isotope and detector---becomes fully evident when considering specific experimental realizations. Three emblematic cases illustrate this point particularly well. The isotope ${}^{76}$Ge can be deployed in large-volume, high-purity germanium diodes~\cite{Fiorini:1960} that combine source and detector in a single element and provide unrivaled energy resolution. This approach underpinned the past success of the pioneering Heidelberg--Moscow~\cite{Baudis:1999} and IGEX~\cite{Aalseth:2002} experiments and was further refined by GERDA~\cite{Agostini:2020} and MAJORANA~\cite{Arnquist:2023}; their merger into the LEGEND program~\cite{Brugnera:2025} now defines the state of the art, demonstrating that excellent background suppression and energy resolution can compensate for a relatively low $Q$-value. In the case of ${}^{130}$Te, large and radiopure TeO$_2$ crystals can be operated as cryogenic bolometers, following the technology proposed in Ref.~\cite{Fiorini:1983}, with excellent energy resolution and scalability. The high natural abundance of ${}^{130}$Te enabled the CUORE experiment~\cite{Adams:2022} to deploy a large isotope mass without enrichment. The bolometric technology now forms the basis for next-generation bolometric searches such as CUPID~\cite{Alfonso:2025} and AMoRE~\cite{Alenkov:2015arXiv}, with improved particle identification and the more favorable isotope ${}^{100}$Mo. Finally, ${}^{136}$Xe offers wide versatility: after a pioneering experiment in a proportional chamber~\cite{Bellotti:1989}, it was and is used as a gas or liquid in time-projection chambers with topological discrimination (like in NEXT~\cite{Martin-Albo:2016} and EXO-200~\cite{Anton:2019}, with the planned NEXT-BOLD, NEXT-CRAB~\cite{Byrnes:2023} and nEXO~\cite{Adhikari:2022} experiments as major future milestones) or dissolved in liquid scintillators to reach very large source masses (like in KamLAND-Zen~\cite{Abe:2025}). The relatively low enrichment cost and the high achievable production capacity further strengthen xenon’s appeal. In an ironic twist, the three isotopes ${}^{76}$Ge, ${}^{130}$Te, and ${}^{136}$Xe are among the least favorable of the ``magnificent nine'' in terms of $Q$-value, as clearly shown in Figure~\ref{fig:candidates}. Nevertheless, they currently provide the most stringent constraints on $0\nu2\beta$. This fact underscores the decisive role of the detection technique in achieving ultimate experimental sensitivity.

    It may appear surprising that NMEs are not listed as a primary criterion for isotope selection. The reason is not that this parameter is unimportant, but rather that the present uncertainties in nuclear-structure calculations (see the discussion in Section~\ref{sec:NME} and Figure~\ref{fig:NME-spread}) prevent NMEs from being used as a reliable indicator of a particularly favorable isotope, apart from the general consensus that ${}^{48}$Ca occupies an unfavorable position. By contrast, the phase-space factor, which is driven by the $Q$-value, is currently a much more robust and reliable quantity. Nevertheless, the existence of a “super-isotope” with exceptionally favorable NMEs cannot be excluded; currently, however, we lack the theoretical tools needed to identify it unambiguously.

\begin{figure}[H]
\centering
      \includegraphics[width=0.9\linewidth]{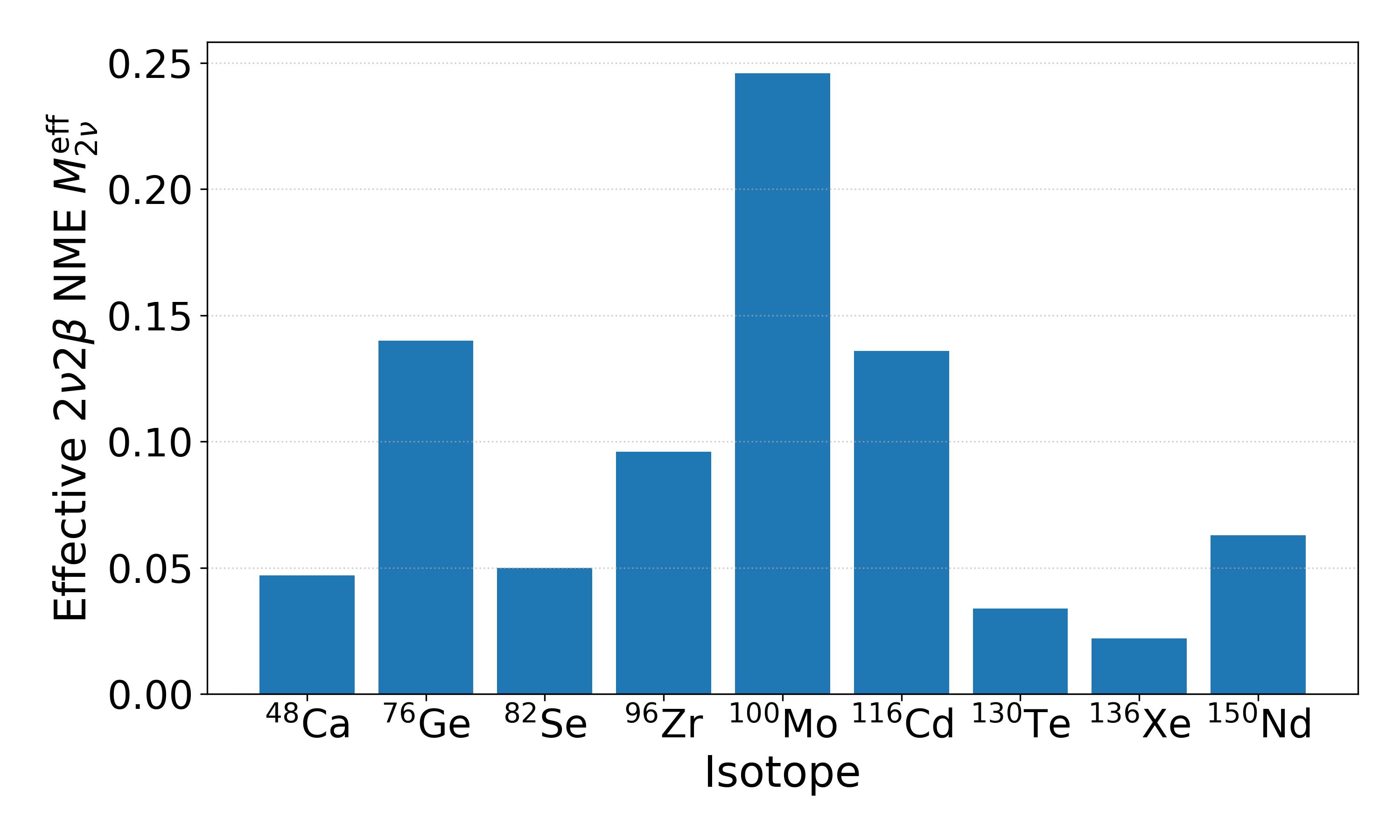}
\vspace{-8pt}
      \caption{Effective $2\nu2\beta$ nuclear matrix elements for the ``magnificent nine" isotopes.}
      \label{fig:2n-NME}
    \vspace{-10pt}
    \end{figure}

    In this respect, it is instructive to consider the case of $2\nu2\beta$ decay, for which experimental measurements of the decay rate exist for several isotopes~\cite{Barabash:2020} and the phase space can be calculated with high precision~\cite{Kotila:2012}, as in the $0\nu2\beta$ case. This makes it possible to extract effective nuclear matrix elements~\cite{Barabash:2020}, $M_{2\nu}^{\rm eff}$, without relying on uncertain nuclear-structure calculations (see Table~\ref{tab:nme}). Applying this procedure, one clearly observes that the $M_{2\nu}^{\rm eff}$ values span a wide range (as shown in Figure~\ref{fig:2n-NME}) and that a true “super-isotope” does exist in this channel, namely $^{100}$Mo, which is characterized by an unusually short $2\nu2\beta$ half-life and high $M_{2\nu}^{\rm eff}$. Of course, there is no guarantee that a similar situation holds for $0\nu2\beta$ decay, nor that an isotope with favorable NMEs in $2\nu2\beta$ retains this feature in the neutrinoless mode. (In addition, this feature may make $^{100}$Mo challenging to use in low time-resolution detection technologies due to accidental coincidences of $2\nu2\beta$ events.) Nevertheless, these observations provide clear evidence that nuclear-structure effects play a significant role. The present situation therefore motivates the exploration of several isotopes with the highest achievable sensitivities, in order to maximize the probability of detecting $0\nu2\beta$.

\vspace{-6pt}

 	\begin{table}[H]
\centering
      \caption{$Q$-values, measured $2\nu2\beta$ half-lives, and effective $2\nu2\beta$ NME $M_{2\nu}^{\rm eff}$ (dimensionless) for the ``magnificent nine" isotopes.} 
      \label{tab:nme}
   \newcolumntype{c}{>{\centering\arraybackslash}X}
\begin{tabularx}{\textwidth}{cccc} 
        \toprule
       \textbf{Isotope} & $\bm{Q}$ \textbf{[keV]} & $\bm{T_{1/2}^{2\nu}}$ \textbf{[yr]} & $\bm{M_{2\nu}^{\rm eff}}$ \\
       \midrule
        $^{48}$Ca  & 4272 & $4.4\times 10^{19}$ & 0.047 \\
        $^{76}$Ge  & 2039 & $1.5\times 10^{21}$ & 0.140 \\
        $^{82}$Se  & 2995 & $9.4\times 10^{19}$ & 0.050 \\
        $^{96}$Zr  & 3350 & $2.3\times 10^{19}$ & 0.096 \\
        $^{100}$Mo & 3034 & $7.1\times 10^{18}$ & 0.246 \\
        $^{116}$Cd & 2813 & $2.8\times 10^{19}$ & 0.136 \\
        $^{130}$Te & 2527 & $6.8\times 10^{20}$ & 0.034 \\
        $^{136}$Xe & 2458 & $2.0\times 10^{21}$ & 0.022 \\
        $^{150}$Nd & 3371 & $8.2\times 10^{18}$ & 0.063 \\
        \bottomrule
      \end{tabularx}
    	\end{table}

\vspace*{-32pt}

     \subsection{The Size of the Challenge}

\begin{figure}[H]
    \centering
    \includegraphics[width=0.8\textwidth]{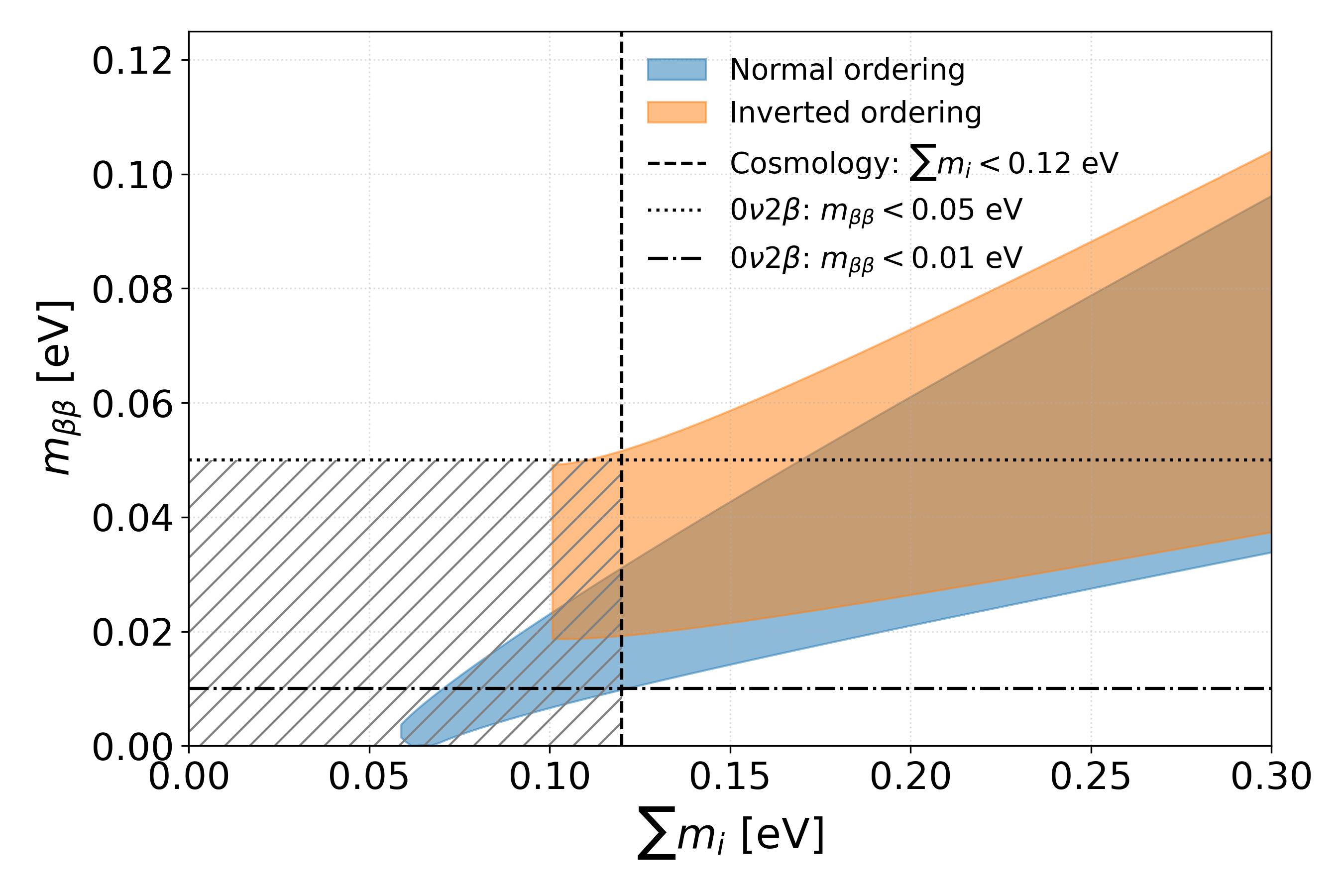}
\vspace{-6pt}
    \caption{The effective Majorana mass $m_{\beta\beta}$ as a function of the sum of the three neutrino masses $\sum m_i$. The allowed regions corresponding to the normal and inverted neutrino mass orderings are shown in blue and orange, respectively. Current constraints from cosmology and $0\nu2\beta$ searches are also indicated. The hatched rectangle approximately highlights the region of parameter space that remains to be explored.}
    \label{fig:lobster}
    \vspace{-18pt}
    \end{figure}

    In order to design a $0\nu2\beta$ experiment, one must first define a clear physics goal, then translate it into a target sensitivity on an experimental observable, and finally develop an experimental setup based on an appropriate technology capable of achieving the required sensitivity. The physics goal is commonly expressed in terms of the effective Majorana mass $m_{\beta\beta}$ within the framework of the mass mechanism and can be conveniently illustrated by Figure~\ref{fig:lobster}. This figure shows the allowed regions of $m_{\beta\beta}$---expressed by Equation~(\ref{eq:mbb})---as a function of the sum of the three neutrino masses $\sum m_i$. These regions are obtained by varying the unknown Majorana phases (see Equation~(\ref{eq:majorana_phases})) over the full interval $[0,2\pi]$ and by adopting the current best-fit values of the leptonic mixing parameters~\cite{Esteban:2020,deSalas:2021,Capozzi:2021}. The shaded bands correspond to the normal (NO) and inverted (IO) neutrino mass orderings and are drawn with semi-transparency to emphasize their overlap in the quasi-degenerate regime. At low values of $\sum m_i$, the NO region extends down to vanishing $m_{\beta\beta}$ due to possible destructive interference among the contributing terms, whereas the IO region exhibits a non-zero lower bound, reflecting the characteristic mass-splitting pattern of this ordering.

    The vertical line indicates the approximate constraint from cosmological observations on $\sum m_i$~\cite{Aghanim:2020}. Recent cosmological analyses have pushed the upper limit on the sum of the three neutrino masses to values that approach, or even undercut, the minimum mass scale implied by oscillation data. Within the standard $\Lambda$CDM model, combinations of cosmic microwave background observations, baryon acoustic oscillations, and other large-scale structure probes have yielded upper limits on $\sum m_i$ at the level of $\sim\!0.05$--$0.07\,\text{eV}$ for some data combinations~\cite{Wang:2024}. These values are close to the lower bound of $\sim\!0.06\,\text{eV}$ required by oscillation experiments for NO, and well below the $\sim\!0.10\,\text{eV}$ minimal value associated with IO. If taken at face value, such stringent cosmological bounds would therefore be in tension not only with the IO but also with expectations for the NO, suggesting a possible inconsistency between cosmological inferences and terrestrial oscillation measurements.
    However, cosmological constraints on neutrino masses are model dependent and sensitive to assumptions about the underlying cosmological framework, the choice of data sets, and possible systematic effects. For this reason, and in order to avoid over-interpreting potentially model-driven results, we adopt a conservative approach and show only the more robust Planck bound $\sum m_i < 0.12\,\text{eV}$ at 95\,\%~C.L. as the cosmological constraint in this plot~\cite{Aghanim:2020}.
    
    Current experiments, described in Section~\ref{sec:status}, have achieved sensitivities to the effective Majorana mass at the level of $\sim\!50\,\text{meV}$, corresponding to the upper horizontal reference line shown in Figure~\ref{fig:lobster}. This limit is significantly affected by uncertainties in NME calculations (see Section~\ref{sec:NME} and Figure~\ref{fig:NME-spread}) and should therefore be regarded as a rough indication. The region of parameter space that remains to be explored is thus represented by the hatched rectangle in Figure~\ref{fig:lobster}. The ambitious goal of next-generation $0\nu2\beta$ experiments is to push the sensitivity on $m_{\beta\beta}$ down to $\sim\!10$~meV, at least for the most favorable NME calculations. \textls[-15]{In order to translate this physics goal into an experimental target and to approximately quantify the associated challenge, one may use the following working formula:}
    \begin{equation}
            T_{1/2}^{0\nu} \sim [10^{27} - 10^{28}] \; 
            \text{yr} \; 
            \left(\frac{0.01~\text{eV}}{m_{\beta\beta}}\right)^2
            \label{eq:approximate-rate}
    \end{equation}
    which is based on Equation~(\ref{eq:rate}) and provides an estimate of the half-life sensitivity required to probe a given value of the effective Majorana mass. This expression is applicable to most of the ``magnificent nine'' candidate isotopes, and the span of one order of magnitude in the prefactor reflects uncertainties in the NMEs as well as the spread in PSFs due to the different $Q$-values and charges of the various isotopes, which were extensively discussed in Sections~\ref{sec:NME} and \ref{sec:phase-space}, respectively. As an example, a value closer to $10^{28}\,\text{yr}$ should be adopted for $^{76}$Ge, which is penalized by its relatively low $Q$-value and, consequently, a smaller PSF. In contrast, a value closer to $10^{27}\,\text{yr}$ can be considered for $^{150}$Nd, owing to its exceptionally large PSF, or for $^{100}$Mo, at least when QRPA and EDF-GCM nuclear models are employed. The candidates $^{136}$Xe and $^{130}$Te are in a somewhat intermediate position, with a large spread in the NME values. 
\vspace{3pt}

    Setting the physics target at $m_{\beta\beta} \simeq 10\,\text{meV}$, Equation~(\ref{eq:approximate-rate}) indicates that half-life sensitivities in the range of $10^{27}$--$10^{28}\,\text{yr}$ are required. The expected number of signal events for the ``magnificent nine'' isotopes, assuming one ton of isotope and an exposure of 10 years, is summarized in Table~\ref{tab:magnificent_nine}. From these estimates, a useful rule of thumb emerges: one expects $\mathcal{O}(30$--$50)$ events per ton$\cdot$10\,y for a half-life of $10^{27}\,\text{yr}$, and only $\mathcal{O}(3$--$5)$ events per ton$\cdot$10\,y for a half-life of $10^{28}\,\text{yr}$.

    These numbers set the characteristic scale for next-generation experiments and, at the same time, define extremely stringent requirements on the allowable background level. In the regime of very small expected signal counts, even a handful of background events within the region of interest can severely degrade the experimental sensitivity or completely obscure a potential signal. As a consequence, future $0\nu2\beta$ experiments are effectively compelled to operate in a near zero-background regime over the entire exposure.
    
    This requirement can be made more quantitative by considering the dependence of the half-life sensitivity on exposure and background. In the background-free limit, the sensitivity scales linearly with exposure,
    \begin{equation}
    S(T_{1/2}^{0\nu}) \propto \varepsilon \, M \, t ,
    \label{eq:zero-background}
    \end{equation}
    where $\varepsilon$ is the detection efficiency, and $M$ and $t$ are the isotope mass and live time, respectively. In contrast, in the background-dominated regime the sensitivity improves only with the square root of the exposure,
    \begin{equation}
    S(T_{1/2}^{0\nu}) \propto \varepsilon \, \sqrt{\frac{M \, t}{B \, \Delta E_{ROI}}},
    \label{eq:nonzero-background}
    \end{equation}
    where $B$ is the background index (in counts/(keV$\cdot$kg$\cdot$y)) and $\Delta E_{ROI}$ is the width of the energy ROI, chosen of the same order as the detector energy resolution $\Delta E_{FWHM}$.

\newpage
    \vspace*{-36pt}
 \begin{table}[H]
    \centering
    \caption{Expected strength of the $0\nu2\beta$ signal for the ``magnificent nine'' candidate isotopes. The second column reports the number of candidate nuclei contained in one metric ton of isotope, assuming 100\,\% isotopic enrichment. The last two columns list the expected number of $0\nu2\beta$ events for an exposure of 10 years, assuming half-lives of $10^{27}$ and $10^{28}$~yr, respectively.}
    \label{tab:magnificent_nine}
    \newcolumntype{c}{>{\centering\arraybackslash}X}
\begin{tabularx}{\textwidth}{cccc} 
    \toprule
    \textbf{Isotope} 
    & $\bm{N}$ \textbf{Nuclei/ton} 
    & $\bm{N_{0\nu}}$ \textbf{(10 yr,} $\bm{10^{27}}$ \textbf{yr)} 
    & $\bm{N_{0\nu}}$ \textbf{(10 yr,} $\bm{10^{28}}$ \textbf{yr)} \\
    \midrule
    $^{48}$Ca   & $1.25\times10^{28}$ & $87$ & $8.7$ \\
    $^{76}$Ge   & $7.9\times10^{27}$  & $55$ & $5.5$ \\
    $^{82}$Se   & $7.3\times10^{27}$  & $51$ & $5.1$ \\
    $^{96}$Zr   & $6.3\times10^{27}$  & $44$ & $4.4$ \\
    $^{100}$Mo  & $6.0\times10^{27}$  & $42$ & $4.2$ \\
    $^{116}$Cd  & $5.2\times10^{27}$  & $36$ & $3.6$ \\
    $^{130}$Te  & $4.6\times10^{27}$  & $32$ & $3.2$ \\
    $^{136}$Xe  & $4.4\times10^{27}$  & $31$ & $3.1$ \\
    $^{150}$Nd  & $4.0\times10^{27}$  & $28$ & $2.8$ \\
   \bottomrule
		\end{tabularx}
    \end{table}

     The zero background regime occurs when the expected background level in the ROI is well below one count over the full dataset, corresponding to the condition $M \, t \, B \, \Delta E _{ROI} \ll 1 $.This illustrates why aggressive background suppression, excellent energy resolution, and high signal efficiency are not merely desirable features, but rather fundamental prerequisites for the next generation of $0\nu2\beta$ experiments.

     \section{Present and Future Experiments}
     \label{sec:experiments}
    
     \subsection{Experimental Approaches}
    
    The extreme rarity of $0\nu2\beta$, with expected half-lives extending potentially well beyond $10^{26}\,\text{yr}$, imposes a set of stringent and often competing experimental requirements. These requirements naturally define the strategies adopted by current and next-generation experiments.
    
    A primary consideration is the strength of the source. Probing half-lives at $10^{25}$-$10^{26}\,\text{yr}$ (as in the current experiments) or at the level of, or beyond, $10^{27}\,\text{yr}$ (as in the next-generation searches) requires an extraordinarily large number of candidate nuclei. This translates into isotope masses at the 100 kg--1000 kg scale, assuming enrichment levels close to 100\,\% (see Table~\ref{tab:magnificent_nine}) . Consequently, all experiments converge today toward the ``source = detector'' paradigm, in which the double beta decay isotope is embedded directly within the active detector medium. While the alternative approach of a spatially separated source and detector offers powerful capabilities for event reconstruction and studies of the underlying decay mechanism, its intrinsically lower efficiency renders it unsuitable for ultimate sensitivity searches. Nevertheless, this technique remains valuable in the event of a discovery, where detailed event topology can help discriminate among competing $0\nu2\beta$ mechanisms. This approach, in fact, enables the determination of additional signatures, including single-electron energy spectra and angular correlations between the two electrons, as demonstrated by the NEMO-3~\cite{Brudanin:2011} experiment with several isotopes and under further investigation with $^{82}$Se in the SuperNEMO demonstrator in the Modane underground laboratory in France~\cite{Aguerre:2025}.
    
    The discussion at the end of the previous section shows that a crucial experimental challenge consists of the background suppression, which becomes increasingly critical as exposures approach the ton$\cdot$year scale.
    Many mitigation strategies are common to rare-event experiments, including underground operation to suppress cosmic-ray–induced backgrounds, massive passive shielding, active veto systems, and the use of ultra-radiopure materials. 
    
\begin{figure}[H]
    \centering
    \includegraphics[width=0.9\textwidth]{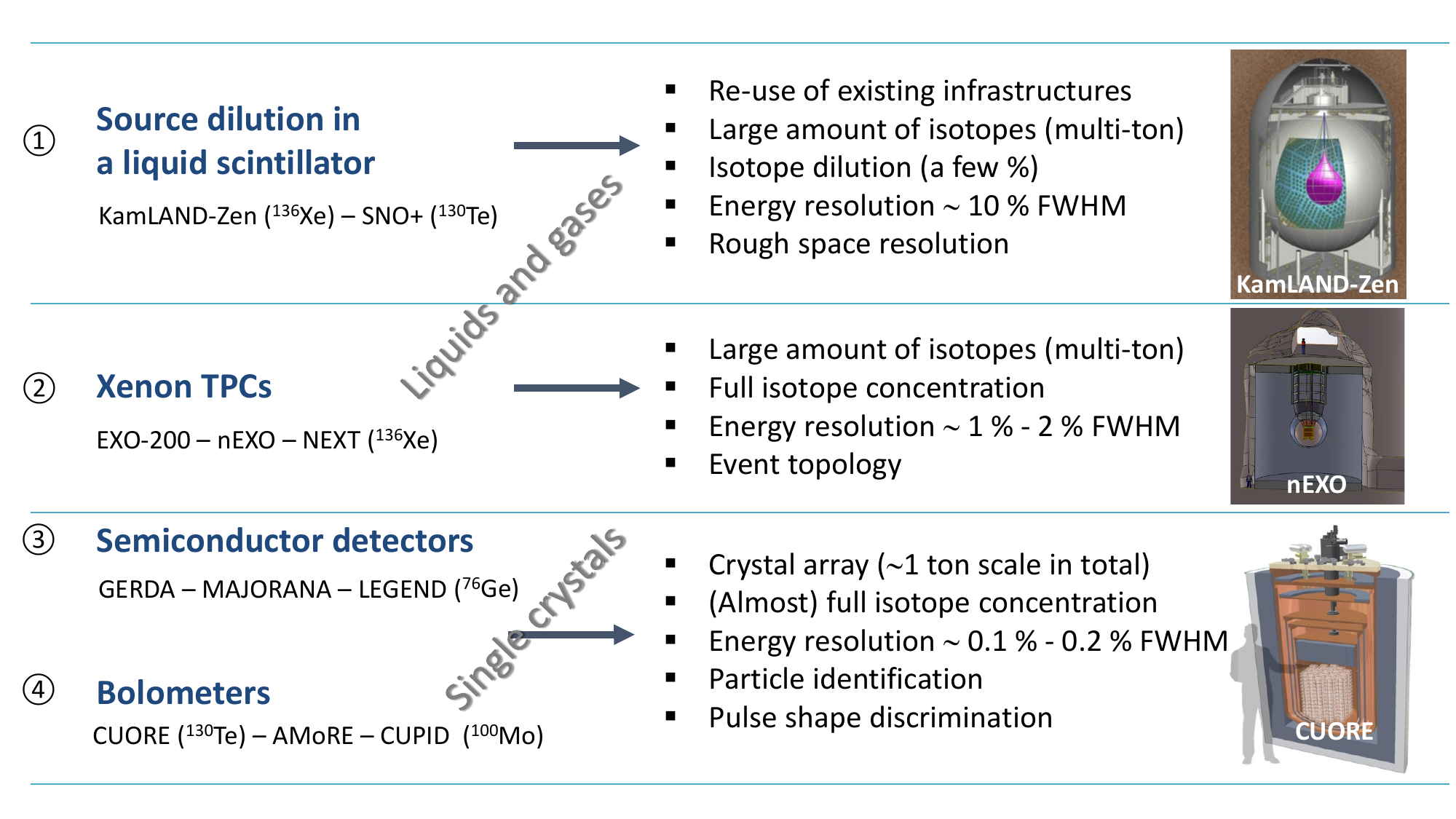}
\vspace{-6pt}
    \caption{\textls[-20]{Classification of ``source = detector'' technologies, showing examples of the most relevant experiments in each category. Distinctive features of each technology are concisely reported, highlighting differences in detector properties.}}
    \label{fig:technologies}
    \vspace{-12pt}
    \end{figure}

   \textls[-15]{ In addition, $0\nu2\beta$ searches impose highly specific requirements driven by the need to isolate a monochromatic peak at the decay $Q$-value. Of course, among the most critical detector parameters is energy resolution. A narrow energy response minimizes the width of the region of interest and suppresses continuum backgrounds (Equations~(\ref{eq:zero-background}) and (\ref{eq:nonzero-background})), including the irreducible tail of $2\nu2\beta$ spectrum (see Section~\ref{sec:signatures}). Particle identification capabilities provide another powerful handle, allowing discrimination between signal-like $\beta$ events and backgrounds from $\alpha$ or surface contaminations. Techniques such as pulse-shape discrimination, dual readout of heat and scintillation light, or correlations between ionization and scintillation are widely employed. Event topology further enhances background rejection. The ability to distinguish single-site from multi-site energy depositions suppresses backgrounds from multi-Compton $\gamma$ interactions, while fiducialization and active shielding reduce surface-related and external backgrounds. In some detector concepts, the identification of the final-state nucleus or daughter ion offers an additional, potentially background-free signature, albeit at the cost of significant technical complexity. Several detector technologies have emerged as leading candidates for $0\nu2\beta$ searches, each representing a different optimization of isotope mass, energy resolution, background rejection, and scalability. The experimental scenario is schematically illustrated in Figure~\ref{fig:technologies}.} 
\vspace{-3pt}
    
    Experiments based on isotope dilution in large liquid scintillator detectors, such as KamLAND-Zen~\cite{Abe:2025} (focusing on $^{136}$Xe) and SNO+~\cite{Albanese:2021} (focusing on $^{130}$Te), benefit from the reuse of existing infrastructures---initially conceived for the study of reactor and solar neutrinos---and the ability to host multi-ton quantities of isotope. Their main limitations arise from relatively modest energy resolution, typically at the level of $\sim\!10\,\%$ FWHM, and limited spatial resolution.

    Time projection chambers (TPCs), particularly those employing high-pressure-gaseous or liquid xenon, provide a powerful compromise between $^{136}$Xe isotope mass and detector performance. They offer large, fully concentrated isotope masses, energy resolution at the percent level, and access to event topology. This class includes experiments such as EXO-200~\cite{Anton:2019}, the proposed nEXO~\cite{Adhikari:2022}, NEXT in its various stages~\cite{Martin-Albo:2016,Byrnes:2023}, and PANDAX-4T~\cite{Yuan:2025arXiv}.

    Semiconductor detectors, exemplified by germanium-based experiments for the study of $^{76}$Ge, achieve exceptional energy resolution at the level of $\sim\!0.1$--$0.2\,\%$ FWHM and operate with fully enriched material. Experiments such as GERDA~\cite{Agostini:2020}, MAJORANA~\cite{Arnquist:2023}, LEGEND~\cite{Brugnera:2025} and CDEX-1B~\cite{Zhang:2024} exploit these strengths to reach some of the lowest background levels achieved in rare-event searches.

    Cryogenic bolometers combine excellent energy resolution (at the level of $\sim\!0.2\,\%$ FWHM) with high isotope concentration and, when operated as scintillating~\cite{Armengaud:2020} or Cherenkov~\cite{Berge:2018} detectors, powerful particle identification capabilities. Large crystal arrays such as CUORE~\cite{Adams:2022} (focusing on $^{130}$Te), CUPID-0~\cite{Azzolini:2022} (focusing on $^{82}$Se), AMoRE~\cite{Alenkov:2015arXiv} and CUPID-Mo~\cite{Augier:2022} (both focusing on $^{100}$Mo) demonstrate the scalability of this approach to the ton scale and its suitability for next-generation $0\nu2\beta$ experiments such as CUPID~\cite{Alfonso:2025} and CUPID-1T~\cite{Armatol:2022arXiv}, which aim to study $^{100}$Mo with unprecedented sensitivity.

    Summarizing, four main technology categories can be identified in the ``source = detector'' approach: liquid-scintillator experiments with dissolved isotopes; xenon TPCs; semiconductor detectors; and cryogenic bolometers. The first two employ a liquid or gaseous medium, with scalability primarily limited by the vessel size, while the latter rely on single crystals and increase total detector mass by assembling larger arrays of identical module (Figure~\ref{fig:technologies}). These technologies are not merely competing but fundamentally complementary, providing cross-checks and robustness against unknown systematics and uncertainties in NME calculations.

    \subsection{Current Experimental Status}
    \label{sec:status}

    Over the past decade, $0\nu2\beta$ searches have attained remarkable progress. The achieved results now form what may be described as a ``restricted club'' of experiments probing half-lives above $10^{24}\,\text{yr}$ for five selected isotopes, belonging to the ``magnificent nine" set: $^{136}$Xe, $^{76}$Ge, $^{130}$Te, $^{82}$Se and $^{100}$Mo. Among the most stringent limits are those obtained by KamLAND-Zen~800~\cite{Abe:2025} on $^{136}$Xe, GERDA~\cite{Agostini:2020} and the MAJORANA Demonstrator on $^{76}$Ge~\cite{Arnquist:2023}, and CUORE~\cite{Campani:2025} on $^{130}$Te. Additional important contributions come from EXO-200~\cite{Anton:2019}, CUPID-0~\cite{Azzolini:2022}, CUPID-Mo~\cite{Augier:2022}, NEMO-3~\cite{Arnold:2015}, and, more recently, AMoRE-I~\cite{Agrawal:2025}, with bolometric experiments playing an increasingly prominent role. At present, most of these experiments have completed data taking, with CUORE representing a notable exception. The most stringent bounds are summarized in Table~\ref{tab:status}. 

\vspace{-12pt}

    \begin{table}[H]
    \centering
    \caption{Summary of current $0\nu2\beta$ limits (at 90\% C.L. or C.I.) achieved by the most advanced past and current experiments. NEMO-3 is the only search in which the source and the detector are separated. The quoted ranges for the effective Majorana mass $m_{\beta\beta}$ reflect the spread of NME calculations and correspond to the choices adopted by each collaboration. This accounts for apparent inconsistencies in the reported $m_{\beta\beta}$ ranges for the same isotope across different experiments. The axial coupling constant $g_A$ is always taken at its free-nucleon value. The exposure corresponds to isotope mass.}
    \label{tab:status}
\newcolumntype{C}[1]{>{\centering\arraybackslash}p{#1}}
\begin{tabular*}{\textwidth}{@{\extracolsep{\fill}} C{4cm} C{1cm} C{3cm} C{2.5cm} C{3cm}} 
   \toprule
    \textbf{Experiment} & \textbf{Isotope} & \textbf{Exposure [kg}$\bm{\cdot}$\textbf{yr]} & $\bm{T_{1/2}^{0\nu}}$ \textbf{Limit [y]} & $\bm{m_{\beta\beta}}$ \textbf{Range [meV]} \\
    \midrule
    KamLAND-Zen 800       & $^{136}$Xe & $\sim\!2100$   & $>3.8\times10^{26}$ & $28$--$122$ \\
    CUORE                 & $^{130}$Te & $372$       & $>3.4\times10^{25}$ & $70$--$250$ \\
    GERDA                 & $^{76}$Ge  & $127$       & $>1.8\times10^{26}$ & $79$--$180$ \\
    EXO-200               & $^{136}$Xe & $234$       & $>3.5\times10^{25}$ & $93$--$286$ \\
    MAJORANA Demonstrator & $^{76}$Ge  & $\sim\!86$        & $>8.3\times10^{25}$ & $113$--$269$ \\
    AMoRE-I               & $^{100}$Mo & $3.89$       & $>2.9\times10^{24}$ & $210$--$610$ \\
    CUPID-0               & $^{82}$Se  & $8.82$        & $>4.6\times10^{24}$ & $263$--$545$ \\
    CUPID-Mo              & $^{100}$Mo & $1.47$       & $>1.8\times10^{24}$ & $280$--$490$ \\
    NEMO-3                & $^{100}$Mo & $34.3$      & $>1.1\times10^{24}$ & $330$--$620$ \\
    \bottomrule
   		\end{tabular*}
    \end{table}

\begin{figure}[H]
    \centering
    \includegraphics[width=1\textwidth]{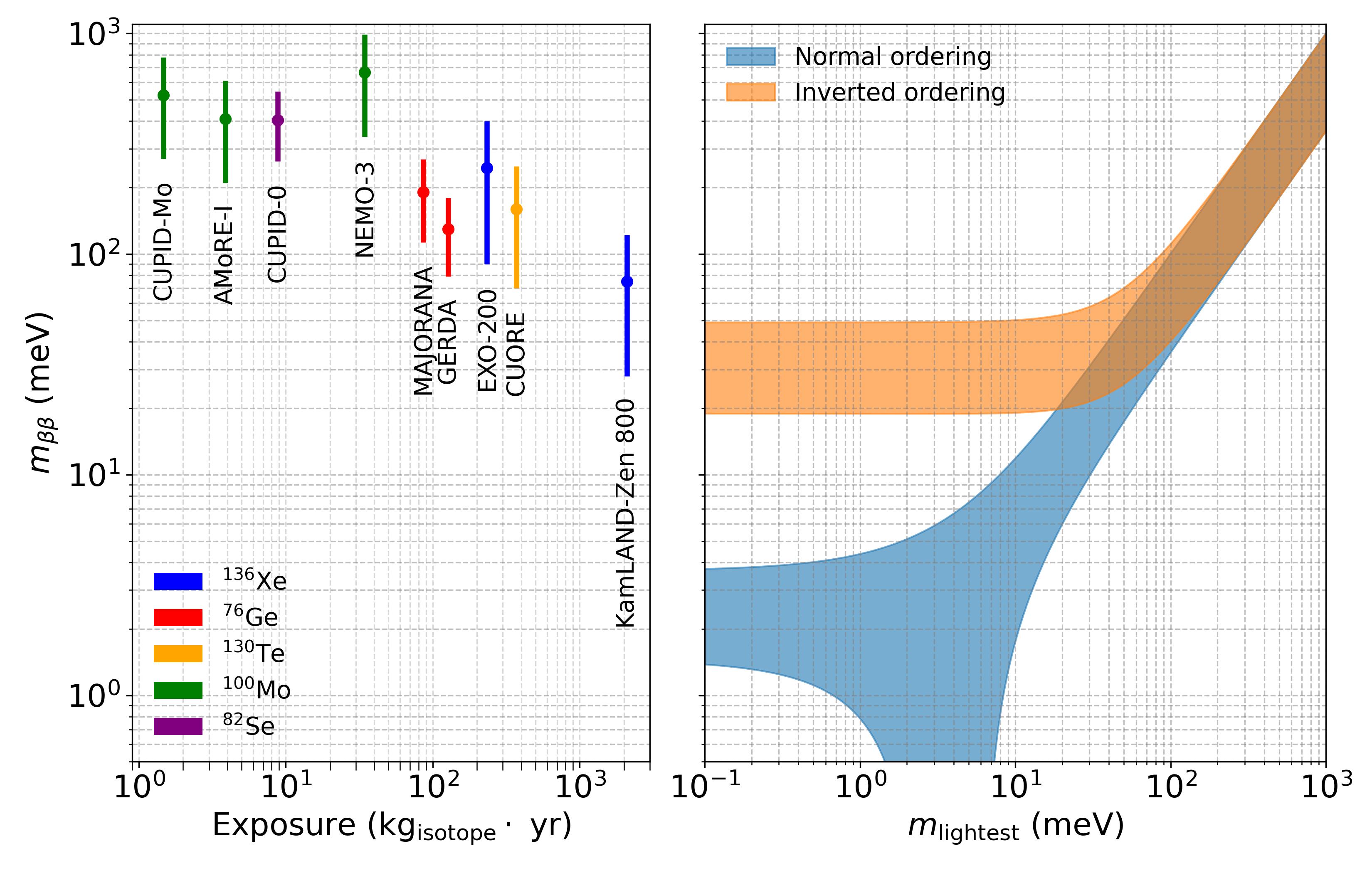}\vspace{-6pt}
    \caption{Experimental status of the search for $0\nu2\beta$. On the left, the ranges of limits on the effective Majorana mass for the experiments listed in Table~\ref{tab:status} are shown as a function of the exposure. On the right, a plot of the effective Majorana mass as a function of the lightest neutrino mass illustrates the location of current experimental bounds within the neutrino mass parameter space, providing an estimate of their physics reach. This plot conveys the same information as Figure~\ref{fig:lobster}. Unlike Table~\ref{tab:status}, the same NMEs are adopted for different experiments studying the same isotope, following the choice used by the most recent experiment.}
    \label{fig:exposure-vs-mbb-status}
    \end{figure}

    Although no signal has yet been observed, the collective achievements of these experiments have established the experimental techniques, background suppression strategies, and scalability required for the next generation of detectors. The most sensitive among them, and especially the leading KamLAND-Zen 800 project, fix the rough current bound of 50~meV shown in Figure~\ref{fig:lobster}, considering the large spread due to NME calculations. The physics reach of the current experiments can be appreciated in Figure~\ref{fig:exposure-vs-mbb-status}.

     \subsection{Planned Projects}
     \label{sec:future}

     A new class of $0\nu2\beta$ experiments is being prepared with the goal of fully exploring the IO region of neutrino masses and, in some cases, extending sensitivity beyond it, at the level of $\mathcal{O}(10\,\text{meV})$. Often, these projects build directly on the technologies, infrastructures, and operational experience of their respective precursor experiments, while introducing substantial improvements in isotope mass, background suppression, and detector performance. The following review of the main searches is not exhaustive, but focuses on the experiments that are most relevant and representative of each technology. The physics reach of an experiment is commonly expressed in terms of its sensitivity to the $0\nu2\beta$ half-life. However, there is no uniform convention for quoting this quantity across collaborations. In some cases, only the ``exclusion sensitivity" is reported, typically at the 90\% confidence level (C.L.) or confidence interval (C.I.). Other collaborations also report---or report exclusively---the more conservative ``discovery sensitivity". This quantity represents the probability (often 50\% or $3\sigma$) that the experiment will observe a signal at or above a specified significance level, assuming the true half-life is equal to the quoted discovery sensitivity value. It is therefore not always possible to compare precisely different experiments on the basis of their quoted sensitivity. However, this limitation is not very relevant when considering the underlying physics parameter, $m_{\beta\beta}$, given that the dominant uncertainty arises from the NMEs.

    The SNO$^+$ experiment~\cite{Albanese:2021}, located underground in SNOLab in Canada, represents one of the earliest large-scale attempts to search for $0\nu2\beta$ decay using an isotope-loaded liquid scintillator, exploiting the existing SNO detector infrastructure used for the detection of solar neutrinos. After an extended commissioning and low-background operation phase, SNO$^+$ has progressed toward its $0\nu2\beta$ program based on the loading of natural tellurium into the scintillator, exploiting, like CUORE,  the uniquely high natural isotopic abundance (34.1~\%) of the double beta decay candidate $^{130}$Te (see Figure~\ref{fig:candidates}). It is worth noting that the start of the $0\nu2\beta$ phase experienced a significant delay with respect to the original schedule: SNO$^+$ was initially expected to begin isotope-loaded operation in the same time frame as CUORE, between roughly 2015 and 2020, and using the same isotope. SNO$^+$ foresees two phases of $0\nu2\beta$ data taking. In the initial phase, 3.9 tons of natural tellurium (0.5\% by mass) will be dissolved in the 780 tons of liquid scintillator. The large detector volume and the high radiopurity achievable with liquid scintillator are expected to result in very low background levels, of order 10 counts/yr in the ROI, dominated by interactions of $^8$B solar neutrinos. Assuming three years of data taking, this phase is expected to reach a half-life sensitivity of $2\times10^{26}$ yr (90\% C.L.), corresponding to an effective Majorana mass range of 29–137 meV. The initial phase is funded, with tellurium loading expected to begin in 2026. A second phase, under consideration, would increase the tellurium loading to 1.5\%, leading to a projected sensitivity of $7\times10^{26}$ yr (90\% C.L.), corresponding to $m_{\beta\beta}<17$–73 meV.

    KamLAND2-Zen~\cite{Weerman:2025} is the planned major upgrade of the successful KamLAND-Zen program, which exploits enriched $^{136}$Xe dissolved in liquid scintillator inside a large spherical volume. In the existing detector (located in the Kamioka mine in Japan), a thin, radiopure nylon balloon holds the xenon-loaded scintillator at the detector center, surrounded by approximately 1\,kton of ultra-clean liquid scintillator that acts as an active shield and provides self-shielding against external radioactivity; this internal vessel is the key to achieving an ultra-low background environment and precise fiducialization. KamLAND2-Zen will retain the core KamLAND liquid scintillator technology but introduce a suite of improvements aimed at substantially enhancing energy resolution and background rejection, thereby suppressing the tail of the two-neutrino double-beta decay spectrum and other backgrounds that currently limit sensitivity. These upgrades include a higher light-yield scintillator, high-quantum-efficiency photomultiplier tubes complemented by new light-collecting mirrors to maximize photon collection, and advanced electronics, resulting in an expected energy resolution improvement from about 9\% FWHM to around 4.5\% FWHM at the $Q$-value of $^{136}$Xe $0\nu2\beta$ decay. The collaboration is also developing an improved scintillation balloon film (e.g.\ polyethylene naphthalate) to enable $\alpha$ tagging of surface backgrounds and exploring additional imaging systems to reject multi-site $\gamma$ backgrounds, as well as methods to mitigate long-lived spallation backgrounds. With these enhancements KamLAND2-Zen is designed to host approximately one ton of enriched $^{136}$Xe in the inner balloon and reach a $0\nu2\beta$ half-life sensitivity of about $2\times10^{27}$\,yr after roughly five years of data taking, corresponding to an effective Majorana mass reach of $\sim20$\,meV or better and covering nearly the entire inverted ordering region. Data taking with the upgraded detector is projected to begin toward the end of the current decade, with full physics sensitivity anticipated in the early 2030s.

    The LEGEND~\cite{Brugnera:2025} program represents the natural evolution of the GERDA and MAJORANA Demonstrator experiments, which established high-purity germanium (HPGe) detectors enriched in $^{76}$Ge as one of the most powerful technologies for current $0\nu2\beta$ searches, following the historical precursors Heidelberg-Moscow and IGEX searches. This approach combines an intrinsically excellent energy resolution, of order 0.1--0.2\% FWHM at the $Q$-value, with extremely low intrinsic radioactivity and mature detector fabrication techniques. A key feature of modern HPGe detectors is the use of point-contact geometries, such as BEGe and inverted-coaxial detectors, which enable highly effective pulse-shape discrimination. By exploiting the characteristic time structure of the charge-collection signals, this technique allows single-site energy depositions, as expected for $0\nu2\beta$ decay, to be efficiently separated from multi-site $\gamma$-induced backgrounds, providing powerful event-by-event background rejection without loss of signal efficiency. LEGEND adopts a staged strategy to progressively scale detector mass while maintaining background-free conditions. The first phase, LEGEND-200, currently operating at the Laboratori Nazionali del Gran Sasso in Italy (LNGS), deploys approximately 200~kg of enriched $^{76}$Ge detectors immersed directly in liquid argon. The argon acts simultaneously as cryogenic coolant, passive shielding, and an active veto through the detection of scintillation light produced by background interactions. Together with pulse-shape discrimination and the use of ultra-clean materials, this configuration targets background indices below $10^{-4}$~counts/(keV$\cdot$kg$\cdot$yr), enabling LEGEND-200 to reach half-life sensitivities at the level of $10^{27}$~yr during data taking in the second half of the 2020s. Recently, LEGEND-200 has released the first physics results~\cite{Acharya:2025} which, combined with those from GERDA and the MAJORANA demonstrator, provide the present most stringent bound---$1.9\times10^{26}$ yr (90\% C.L.)---on the $0\nu2\beta$ half-life of $^{76}$Ge. The ultimate goal of the program is LEGEND-1000, a ton-scale experiment based on approximately 1000~kg of enriched $^{76}$Ge, designed to operate in a near background-free regime over exposures of order 10~ton$\cdot$yr. Building on the technologies validated in GERDA, MAJORANA, and LEGEND-200, LEGEND-1000 will further improve background suppression down to $10^{-5}$~counts/(keV$\cdot$kg$\cdot$yr) and below, through enhanced liquid-argon veto performance, optimized detector geometries, stricter material selection, and refined pulse-shape analysis. Under these conditions, LEGEND-1000 is expected to reach a $0\nu2\beta$ half-life sensitivity exceeding $10^{28}$~yr, corresponding to sensitivity to the effective Majorana neutrino mass in the range of 9--21~meV. This sensitivity is sufficient to comprehensively probe the IO region in the neutrino-mass parameter space and extends well into the region of interest for NO scenarios. Construction and commissioning of LEGEND-1000 are foreseen for the early to mid-2030s, with physics data taking expected to follow shortly thereafter.

    The nEXO~\cite{Adhikari:2022} experiment is the envisioned successor to EXO-200 and will significantly advance the TPC technology for $0\nu2\beta$ searches. nEXO will instrument a single, monolithic TPC containing approximately 5\,tons of liquid xenon enriched to about 90\% in $^{136}$Xe. The TPC concept provides simultaneous measurement of both ionization charge and scintillation light, enabling excellent energy resolution through combined charge–light reconstruction, three-dimensional event localization, and powerful discrimination of multi-site gamma and other backgrounds via topological event characterization. The detector design includes a low-radioactivity copper cryostat, ultra-low background materials, a reflective field cage to maximize light collection, and arrays of silicon photomultipliers (SiPMs) optimized for xenon scintillation wavelengths, all immersed in the xenon to achieve high light collection and uniform response. Calibrations using internal and external sources, and sophisticated reconstruction algorithms, are anticipated to yield an energy resolution of about 2.3\% FWHM at the $Q_{\beta\beta}$ value, a key driver for background suppression. With its large target mass and stringent background control, nEXO is projected to reach a $0\nu2\beta$ half-life sensitivity of $1.4\times10^{28}$\,yr (90\% C.L.) after approximately ten years of live data taking, with an effective background index of $7\times10^{-5}$~counts/(keV$\cdot$kg$\cdot$yr) in the optimized analysis region and realistic detector performance. Under typical nuclear matrix element assumptions, this corresponds to an effective Majorana mass reach of 4.7–20.3\,meV, sufficient to cover the entire IO and to begin probing significant regions of NO parameter space. nEXO is currently in an advanced R\&D and engineering design phase, with prototyping of key systems, such as the SiPM arrays and low-background charge readout, well underway. 

    The NEXT~\cite{Martin-Albo:2016,Byrnes:2023} program employs high-pressure xenon gas time projection chambers (HPXeTPCs) with electroluminescent (EL) amplification of ionization to search for $0\nu2\beta$ of $^{136}$Xe. In this technique, charged particles traversing the high-pressure xenon gas produce both prompt scintillation light and ionization electrons. The ionization electrons are drifted by an electric field toward an EL region where they generate proportional scintillation light; the prompt and EL light are detected by separate planes of photosensors, typically photomultiplier tubes (PMTs) for calorimetry and silicon photomultipliers (SiPMs) for tracking. This dual readout enables excellent energy resolution, at or below the 1\% FWHM level in the ROI, together with detailed three-dimensional reconstruction of event topology, which provides powerful discrimination against background events based on track shape and length. The capability to distinguish the characteristic two-electron track topology expected for $0\nu2\beta$ decays from single-electron or multi-site gamma backgrounds is a key feature of the HPXeTPC approach.  Furthermore, the NEXT concept offers scalability to larger detector masses and the prospect of near background-free operation through the implementation of barium tagging~\cite{Byrnes:2024}, which is pursued \textls[-15]{also by the nEXO collaboration~\cite{Chambers:2019}. In a double beta  decay of $^{136}$Xe, the daughter nucleus is produced as a barium ion (initially Ba$^{++}$), which rapidly undergoes charge exchange in the xenon gas to Ba$^{+}$ through interactions with the medium. The identification of this single barium ion in spatial and temporal coincidence with the reconstructed energy deposit and the characteristic two-electron track topology would provide an unambiguous, event-by-event signature of the decay, effectively eliminating all radioactive and cosmogenic backgrounds, making $2\nu2\beta$ the only relevant background source. The NEXT collaboration is pursuing several complementary Ba-tagging techniques based on atomic and molecular physics ~\cite{Byrnes:2024},} including laser-induced fluorescence of Ba$^{+}$ and the use of single-molecule fluorescent chemosensors capable of selectively binding and optically reporting the presence of barium ions in xenon. Proof-of-concept demonstrations of these approaches have already been achieved under relevant experimental conditions, motivating its integration into future large-scale HPXeTPC detectors. After successful prototyping and operation of earlier detectors such as NEXT-White, the NEXT-100 detector---the third stage of the program---has been constructed and commissioned at the Laboratorio Subterráneo de Canfranc (LSC) in Spain and began operation with xenon gas in 2024. NEXT-100 is an asymmetric EL HPXeTPC designed to contain up to approximately 100\,kg of xenon gas enriched to about 90\% in $^{136}$Xe at a pressure of 15\,bar. Its active volume is on the order of a meter in diameter and length, with an energy plane instrumented with an array of PMTs for precise calorimetric measurement and a tracking plane of thousands of SiPMs for detailed reconstruction of ionization trails. The detector is surrounded by passive shielding. Material screening and design optimizations aim for a background index below $\sim10^{-3}$\,counts/(keV$\cdot$kg$\cdot$yr) in the region of interest. With three years of effective data taking, NEXT-100 is expected to reach a $0\nu2\beta$ half-life sensitivity at the level of $6\times10^{25}$\,yr (90\% C.L.), yielding a competitive probe of the $m_{\beta\beta}$ parameter space with current HPXeTPC technology. The commissioning of NEXT-100 has demonstrated stable operation and energy resolution consistent with design expectations. Looking beyond NEXT-100, the collaboration is preparing for ton-scale detectors such as NEXT-HD, which would increase the xenon mass by an order of magnitude and target sensitivities above $10^{27}$\,yr within a few years of operation, and the NEXT-BOLD program, which aims to integrate barium ion tagging capabilities that could reduce backgrounds to nearly zero and enable sensitivities approaching or exceeding $10^{28}$\,yr. 

    Cryogenic scintillating bolometers represent one of the most mature and powerful technologies for next-generation searches for $0\nu2\beta$~\cite{Poda:2021}. In these detectors, energy deposited by a particle interaction in a crystal operated at millikelvin temperatures is measured through the induced phonon signal, while a simultaneous scintillation light signal is detected by an auxiliary cryogenic light absorber. The combined heat–light readout enables efficient particle identification, in particular strong rejection of $\alpha$-induced backgrounds, which limited earlier purely thermal bolometric experiments like Cuoricino~\cite{Andreotti:2011}, CUORE-0~\cite{Alfonso:2015} and CUORE~\cite{Adams:2022}. Different implementations employ either neutron-transmutation-doped (NTD) Ge thermistors or magnetic metallic calorimeters (MMCs) for phonon readout, reflecting different optimization strategies in terms of scalability, timing, and multiplexing. 
    
    Within this framework, the AMoRE program~\cite{Alenkov:2015arXiv} has explored scintillating bolometers based on molybdate crystals enriched in $^{100}$Mo and MMC phonon readout. The AMoRE-I phase~\cite{Agrawal:2025} (located in the Y2L undeground laboratory in South Korea) demonstrated the feasibility of operating enriched scintillating bolometers underground with competitive energy resolution and background control. Building on this experience, AMoRE-II has evolved from the initial CaMoO$_4$ concept to Li$_2$MoO$_4$ crystals enriched in $^{100}$Mo (validated by LUMINEU~\cite{Armengaud:2017} and CUPID-Mo~\cite{Armengaud:2020} and previously selected by CUPID), benefiting from improved crystal quality, lower intrinsic backgrounds, and superior thermal performance. AMoRE-II aims to deploy an increased isotope mass with enhanced background rejection using a new cryogenic facility located in the Yemi Underground Laboratory in South Korea. It targets half-life sensitivities in the $10^{26}$–$10^{27}$\,yr range on a timescale extending into the next decade. 
    
    CUPID~\cite{Alfonso:2025} constitutes the flagship realization of the scintillating bolometer approach and is conceived as the direct successor of CUORE. CUPID capitalizes on the existing CUORE infrastructure at LNGS---including the large-volume dilution refrigerator, cryostat, shielding, and underground installation---ensuring continuity of expertise while introducing a decisive technological evolution. In contrast to CUORE, which employed TeO$_2$ crystals without particle identification, CUPID will deploy arrays of scintillating bolometers, with Li$_2$MoO$_4$ crystals enriched in $^{100}$Mo as the baseline absorber and NTD Ge thermistor readout. The heat-light readout will suppress $\alpha$ backgrounds by at least three orders of magnitude relative to CUORE. A distinctive challenge for large-mass $^{100}$Mo experiments---especially if using slow NTD Ge thermistors as in CUPID---arises from random coincidences of $2\nu2\beta$ events~\cite{Chernyak:2012}, enhanced by the relatively short $2\nu2\beta$ half-life of $^{100}$Mo (see Table~\ref{tab:nme}). CUPID addresses this through the development of ultrasensitive light detectors exploiting the Neganov–Trofimov–Luke (NTL) amplification effect, which dramatically enhances the signal-to-noise ratio and therefore the pulse-pair resolving time of the light channel. Light detectors adopting these techniques were successfully tested within the CROSS project~\cite{Armatol:2025arXiv}, which has also developed a mid-scale demonstrator for the CUPID technology based on 32 Li$_2$MoO$_4$ crystals enriched in $^{100}$Mo~\cite{Giuliani:2025}.  CUPID will be realized in two phases. CUPID-Stage-I, currently in an advanced state of preparation, will deploy 80\,kg of enriched $^{100}$Mo using fully validated technologies within the existing CUORE cryostat. The target background index is at the level of $\sim10^{-4}$\,counts/(keV$\cdot$kg$\cdot$yr), enabling sensitivities of $2\times10^{26}$\,yr after three years of data taking. Depending on the choice of nuclear matrix elements, this corresponds to an effective Majorana mass sensitivity in the range $m_{\beta\beta}\sim25$–$72$\,meV. CUPID-Stage-I could take a leading role in the global $0\nu2\beta$ search at the beginning of the next decade. The full CUPID experiment will further increase the isotope mass to approximately 240\,kg of enriched $^{100}$Mo with $\sim1600$ scintillating bolometers while approaching background-free conditions over ten years exposure. With a projected half-life sensitivity exceeding $10^{27}$\,yr and a corresponding $m_{\beta\beta}$ reach of about 9.5–28\,meV, CUPID will probe the inverted ordering region comprehensively and extend sensitivity toward the normal ordering parameter space at the end of 2030s.

\begin{figure}[H]
    \centering
    \includegraphics[width=0.8\textwidth]{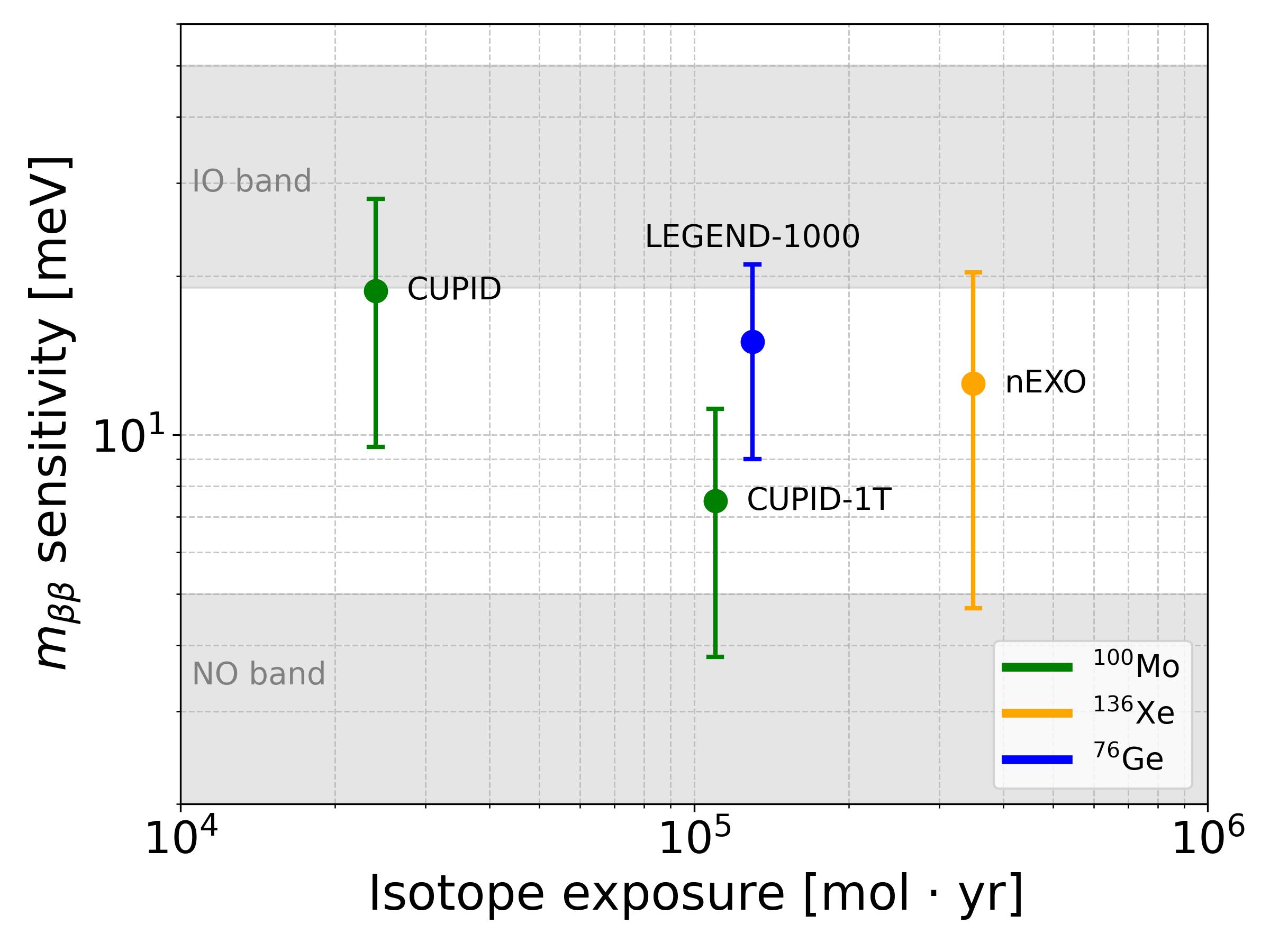}
\vspace{-6pt}
    \caption{Sensitivity of four next-generation $0\nu2\beta$ experiments as a function of the isotope exposure. The experiments---CUPID \& CUPID-1T, LEGEND-1000 and nEXO---are represented with vertical lines indicating the projected range of the bounds on $m_{\beta\beta}$ arising from different NME calculations. The round marker is placed at the center of the range. The IO and NO bands are shown in the limit of vanishing lightest neutrino mass.}
    \label{fig:comparison}
    \vspace{-12pt}
    \end{figure}

    Taken together, all these experiments form a coherent and complementary global program, combining different isotopes and detection techniques, and are expected to decisively test the Majorana nature of neutrinos over the coming decade. In Figure~\ref{fig:comparison}, we compare the sensitivities of the three most advanced experiments in terms of schedule and potential funding: CUPID, LEGEND-1000, and nEXO. Notably, CUPID, which can be realized at a fraction of the cost of the other two projects---primarily due to the existing infrastructure---achieves comparable sensitivities with a much lower exposure, thanks to the favorable properties of the isotope $^{100}$Mo. We also include CUPID-1T in this plot, discussed in the next section, as it represents the final stage of the CUPID sequence approach, with an exposure close to (but still slightly below) that of LEGEND-1000.

     \section{Perspectives toward Sub--10 meV Sensitivity}

    The next generation of $0\nu2\beta$ experiments currently under construction or in advanced planning stages possesses a strong discovery potential in the event that the neutrino mass spectrum follows the IO pattern. In this scenario, the effective Majorana mass $m_{\beta\beta}$ is bounded from below at the level of $\sim20$~meV, and several forthcoming experiments are expected to observe a signal if neutrino is a Majorana particle. If, however, NO holds, the experimental challenge may become more demanding. In this case, $m_{\beta\beta}$ is governed by the lightest neutrino mass $m_1$, and by possible destructive interference effects among the contributions of the three neutrino mass eigenstates. In particular, an accidental relation involving the neutrino masses, the $\theta_{12}$ and $\theta_{13}$ mixing angles and the Majorana phases can lead to exact cancellation and vanishing $m_{\beta\beta}$, as it is apparent in Figures~\ref{fig:lobster} and \ref{fig:exposure-vs-mbb-status} (right panel). As a consequence, the only viable strategy for experimentalists is to pursue ever--increasing sensitivity by extending isotope masses, reducing backgrounds toward the zero--background regime, and developing new and ambitious detector concepts. \textls[-15]{In this context, reaching sensitivity to $m_{\beta\beta} < 10$~meV---corresponding to $T_{1/2}^{0\nu} \sim 10^{28}$--$10^{30}$~yr depending on the  isotopes and the nuclear models (see Equation~(\ref{eq:approximate-rate}))---becomes a critical objective for addressing the NO parameter space.}

       One major and already quite mature direction is the extension of CUPID to the one-ton scale (the CUPID-1T project~\cite{Armatol:2022arXiv}). The baseline solution involves the isotope $^{100}$Mo, in continuity with CUPID-Mo, CUPID, and AMoRE. The increase of the CUPID sensitive mass by a factor of 4 must be accompanied by a reduction of the background index by a factor of 20, down to $5 \times 10^{-6}$~counts/(keV$\cdot$kg$\cdot$yr). Thanks to the detailed, data-driven background model developed for CUPID~\cite{Alfonso:2025}, it is possible to precisely define the detector features required to reduce the background to the targeted level. In CUPID, the two main background sources are random coincidences of $2\nu2\beta$ events and radioactive contamination of nearby detector components made of copper and plastic. The first contribution can be mitigated with advanced fast light detectors using superconducting Transition Edge Sensors, as proposed in Ref.~\cite{Singh:2023} or succesfully implemented in the dark-matter CRESST experiment~\cite{Abdelhameed:2019}, instead of NTD Ge thermistors; the second can be reduced through a new, revolutionary detector arrangement as the one proposed in the BINGO project~\cite{Armatol:2024}, in which the light detector also serves as an active shield against the radioactivity of the passive holder elements. Rather than a radically new project, the $^{100}$Mo-based CUPID-1T can be considered the ultimate stage of CUPID, in natural continuity with CUPID-Stage-I and the full CUPID experiment. Of course, new cryogenic infrastructures would be required, consisting of either a single large cryostat or a distributed multi-cryostat configuration, taking advantage of the major progress in large-volume dilution refrigerators stimulated by developments in superconducting qubits. The targeted half-life sensitivity for CUPID-1T is $T_{1/2} \sim 10^{28}$~yr, corresponding to a sensitivity to $m_{\beta\beta}$ in the range 3.8–11.2\,meV (see Figure~\ref{fig:comparison}).

        Despite the excellent prospects for $^{100}$Mo, the versatility of the bolometric technique allows us to consider other solutions. Another possible candidate is $^{130}$Te, which could become competitive again thanks to new detector developments such as those studied in the BINGO project~\cite{Armatol:2024}, including an inner active veto to mitigate the $\gamma$~background (the ROI for $^{130}$Te lies below 2615~keV, as shown in Figure~\ref{fig:candidates}) and advanced Neganov–Trofimov–Luke light detectors to reject $\alpha$ background via Cherenkov light (TeO$_2$ does not emit appreciable scintillation light)~\cite{Berge:2018}. A potential breakthrough would be the study of the almost background-free isotopes $^{96}$Zr and $^{150}$Nd (Figure~\ref{fig:candidates}), the latter offering a substantial advantage in terms of PSF. Two factors currently limit the use of these candidates: the lack of a scalable enrichment method and the difficulty of identifying an appropriate crystalline compound for bolometric operation. The latter issue is under intensive investigation within the TINY project~\cite{Zolotarova:2024Neutrino}. Advances in enrichment technologies are also possible and would receive strong motivation if bolometers incorporating these materials were successfully developed.
    
        The LEGEND program~\cite{Brugnera:2025}, based on enriched $^{76}$Ge, foresees an ultimate expansion from the current 48 strings of LEGEND-1000 to 133 strings, using larger detectors. The total germanium mass will be approximately 6~tons (LEGEND-6000)~\cite{OrebiGann:2024,Saakyan:2022}. The target background indices are at or below $10^{-6}$~counts/(keV$\cdot$kg$\cdot$yr). Achieving this scale requires roughly a factor of 2 increase in the detector production rate and a factor of 2.5 increase in the isotope production rate. The experiment aims for a background reduction of approximately a factor of 3--5, reaching levels at or below $10^{-6}$~counts/(keV$\cdot$kg$\cdot$yr), together with an order-of-magnitude improvement in half-life sensitivity up to $T_{1/2}\sim10^{29}$~yr, corresponding to an effective Majorana mass sensitivity of $m_{\beta\beta}\sim3$--6~meV. This approach provides a highly complementary pathway to cryogenic bolometric experiments.
    
       Large xenon TPCs represent another promising avenue \cite{Avasthi:2021,OrebiGann:2024}. If scaled to the 0.1--1~kiloton range, they could potentially reach a half-life sensitivity of $T_{1/2}\sim10^{30}$~yr. In their liquid nEXO-style implementation~\cite{Adhikari:2022}, these detectors benefit from intrinsic self-shielding, while the dominant backgrounds arise from $2\nu2\beta$ and $^8$B solar neutrinos. Achieving this sensitivity requires an energy resolution better than or equal to 1\% FWHM and highly efficient use of the target volume, minimizing materials other than the isotope of interest. NEXT-style tracking detectors implemented with high-pressure gas TPCs offer complementary strengths through topological reconstruction~\cite{Byrnes:2023}. They are scalable, with energy resolution better than 1\% FWHM at the double beta decay $Q$-value already demonstrated. Further sensitivity gains may be achievable through barium tagging~\cite{Byrnes:2024,Chambers:2019}, with multiple approaches under development, including ion-to-sensor and sensor-to-ion techniques. The procurement of enriched xenon is itself a significant challenge in this context. Current global xenon production is approximately 50--100~tons per year, largely driven by oxygen production for the steel industry. Additional sources include nuclear fuel processing, which can yield xenon already enriched to about 40\% in $^{136}$Xe. Direct air capture is being explored as an efficient and scalable alternative, with R\&D underway on pilot plants based on advanced adsorbents optimized for xenon extraction.

        A novel approach, with no precursor in current searches, is proposed in the Selena project, which is based on a solid-state, high-resolution imaging detector using enriched $^{82}$Se as both the source isotope and detection medium~\cite{Chavarria:2017,Chavarria:2024}. The detector consists of thin layers of amorphous $^{82}$Se deposited directly onto large-area CMOS active pixel sensors, enabling two-dimensional imaging of ionization tracks with spatial resolutions of order tens of micrometers. This allows direct reconstruction of the distinctive two-electron topology of $0\nu2\beta$ events, providing powerful background rejection through topological and morphological discrimination. An energy resolution of about 1\% FWHM at the $^{82}$Se $Q$-value is sufficient to suppress backgrounds from the high-energy tail of $2\nu2\beta$ decay. A key feature of Selena is its intrinsic scalability to the ton and multi-ton scale, supported by simulations indicating that an exceptionally low background index at the level of $1.5\times10^{-6}$~counts/(keV$\cdot$kg$\cdot$yr) could be within reach of this technology. Each detector module is a self-contained CMOS wafer coated with a millimeter-scale selenium layer, and large target masses are achieved by tiling many identical modules. The use of mature CMOS fabrication and established selenium deposition techniques enables a realistic path toward industrial-scale production with controlled cost and performance. With multi-ton exposures, Selena could reach sensitivity to effective Majorana neutrino masses in the few-meV range, extending well into the normal ordering.

    Another promising avenue is a $0\nu2\beta$ search exploiting the JUNO detector~\cite{Zhao:2017}, following the approach of SNO+ and KamLAND-Zen. JUNO is a 20-kton underground liquid scintillator experiment primarily designed to determine the neutrino mass ordering through precision reactor antineutrino measurements. In the JUNO detector---which features $\sim$4.5\% FWHM energy resolution at the $^{136}$Xe $Q$-value---it is feasible to insert into the central region a balloon filled with enriched-xenon-loaded liquid scintillator. Detailed studies show that with 5 tons of fiducial $^{136}$Xe and 5 years exposure JUNO could reach a sensitivity of $T_{1/2}^{0\nu}\sim5.6\times10^{27}$ yr at 90\% C.L. (and up to $\sim\!1.8\times10^{28}$ yr with $\sim$50 tons), corresponding to effective Majorana mass limits in the few–10 meV range, provided internal and cosmogenic backgrounds are sufficiently controlled through purification and veto strategies. In addition to xenon, natural tellurium dissolved in a similar balloon is also being considered as an alternative isotope.

    Finally, the THEIA concept~\cite{Askins:2020,OrebiGann:2024} represents a hybrid Cherenkov and scintillation detector designed to enhance background rejection through particle identification and event topology reconstruction. THEIA is envisioned as a scalable, ultra-clean liquid detector, with the potential deployment of a 25~kiloton module. It supports a broad physics program, including the study of $0\nu2\beta$ through the insertion of a 16~m radius balloon containing liquid scintillator loaded at the 5\% level with natural tellurium. The projected ten-year sensitivity is $T_{1/2}\sim1.1\times10^{28}$~yr, corresponding to constraints on $m_{\beta\beta}$ extending below 10~meV.

    \textls[-22]{In summary, the convergence of ton-scale isotope masses, ultra-low background levels, and mature detector technologies makes sensitivity to $m_{\beta\beta}<10$~meV an extremely challenging yet feasible goal for mid- to long-term $0\nu2\beta$ searches, with multiple complementary approaches providing robustness against nuclear and experimental uncertainties.}
    
    \section{Conclusions}

    The search for $0\nu2\beta$ remains one of the most powerful and incisive probes of physics beyond the SM. In this review we have summarized the theoretical foundations of the process, its connection to the Majorana nature of neutrinos and lepton number violation, and the role of NMEs in relating experimental observables to fundamental parameters. We have also discussed the present experimental landscape, highlighting the impressive progress achieved over the last two decades in terms of isotope mass, background suppression, energy resolution, and detector scalability, as well as the prospects of the next generation of experiments.
    
    In the coming years, neutrino oscillation experiments are expected to determine the ordering of neutrino masses, providing a crucial piece of information for the interpretation of $0\nu2\beta$ searches. However, oscillation experiments are intrinsically insensitive to the absolute neutrino mass scale and to the Majorana or Dirac nature of neutrinos. Moreover, although cosmological observations currently place stringent constraints on the sum of neutrino masses, the robustness of these bounds depends on the underlying cosmological model and on the treatment of systematic uncertainties, and therefore requires further confirmation.
    
    Beyond laboratory neutrino physics, $0\nu2\beta$ occupies a central position in the broader context of cosmology. Its observation would demonstrate the violation of total lepton number, providing a key experimental ingredient for many theoretical scenarios that aim to explain the observed matter–antimatter asymmetry of the Universe through leptogenesis. While the CP-violating phases accessible in $0\nu2\beta$ are not, in general, sufficient to establish a direct quantitative link to the baryon asymmetry, its observation would nonetheless validate essential prerequisites of these cosmological mechanisms.
    
    In this context, $0\nu2\beta$ plays a unique and irreplaceable role: it is directly sensitive to the absolute neutrino mass scale through the effective Majorana mass, providing information that is complementary to and independent of that obtained from oscillation and cosmological measurements, and with a sensitivity that exceeds the present and foreseeable reach of direct searches based on single beta-decay kinematics. Even in the case of NO, the discovery potential of future planned experiments remains high. Next-generation searches aim to reach sensitivities that cover a significant fraction of the NO parameter space, and more ambitious projects---currently at the conceptual or R\&D level---envision sensitivities extending below the $10\,\mathrm{meV}$ scale. 
    
    Searches for $0\nu2\beta$ will remain a central pillar of neutrino physics and astroparticle physics in the decades to come, with the potential to shed light on the origin of neutrino mass, the violation of fundamental symmetries, and the cosmological origin of the matter–antimatter asymmetry.
 
		\section*{Funding}
This research received no external funding. 
 
		\section*{Data Availability Statement}
Not applicable.
 
		\section*{Conflicts of Interest}
The author declares no conflict of interest.

\section*{Use of AI and AI-Assisted Technologies}
During the preparation of this work, the author used ChatGPT (OpenAI, GPT-5.3), a large language model, to assist with minor language corrections and refinements of the text, as well as the preparation of graphical materials. After using this tool, the author reviewed and edited the content as needed and takes full responsibility for the content of the published article.
	
	\small
	\bibliographystyle{scilight}

	

\end{document}